\documentclass[11pt]{amsart}

\usepackage[colorlinks = true,
            linkcolor = red,
            urlcolor  = magenta,
            citecolor = black,
            anchorcolor = blue]{hyperref}
\usepackage{color}
\usepackage[shortalphabetic,initials,nobysame]{amsrefs}
\usepackage{upgreek}
\usepackage{tikz}
\usepackage[applemac]{inputenc} % for Macs
\usetikzlibrary{arrows}
\usepackage{xcolor}
\usepackage[all]{xy}
\usepackage{graphicx}
\usepackage{wrapfig}
\usepackage{yfonts}
\usepackage{graphicx}
\usepackage{amssymb}
\usepackage{epstopdf}
\usepackage{mathrsfs}
\usepackage[vcentermath]{youngtab}
\usepackage{bm}

\usepackage[mathscr]{euscript}
 \let\mathscr\relax% just so we can load this and rsfs
\usepackage[scr]{rsfso}
\newcommand{\powerset}{\raisebox{.15\baselineskip}{\Large\ensuremath{\wp}}}

\makeatletter \@addtoreset{equation}{section} \makeatother

\renewcommand{\eprint}[1]{\href{https://arxiv.org/abs/#1}{#1}}

\BibSpec{article}{%
    +{}  {\PrintAuthors}                {author}
    +{,} { \textit}                     {title}
     +{,} {}                            {note}
    +{.} { }                            {part}
    +{:} { \textit}                     {subtitle}
    +{,} { \PrintContributions}         {contribution}
    +{.} { \PrintPartials}              {partial}
    +{,} { }                            {journal}
    +{}  { \textbf}                     {volume}
    +{}  { \PrintDatePV}                {date}
    +{,} { \issuetext}                  {number}
    +{,} { \eprintpages}                {pages}
    +{,} { }                            {status}
    +{,} { \DOI}                   {doi}
    +{,} { \eprint}        {eprint}
      +{,} {\publisher}                {publisher}
    +{,} { \address}                   {address}
    +{}  { \parenthesize}               {language}
    +{}  { \PrintTranslation}           {translation}
    +{;} { \PrintReprint}               {reprint}
    +{.} {}                             {transition}
    +{}  {\SentenceSpace \PrintReviews} {review}
}
\def\ee{\end{eqnarray}}
\newcommand{\tp}{\widetilde{\mathfrak{p}}}

\newcommand{\Complex}{\mathbb{C}}

\newcommand{\CN}{\mathcal{N}}

\makeatletter
\newcommand{\ellSN}{\mathop{\operator@font sn}\nolimits}
\newcommand{\ellCN}{\mathop{\operator@font cn}\nolimits}
\newcommand{\ellDN}{\mathop{\operator@font dn}\nolimits}
\newcommand{\ellAM}{\mathop{\operator@font am}\nolimits}
\newcommand{\ellK}{\mathop{\smash{\operator@font K}\vphantom{a}}\nolimits}
\newcommand{\ellE}{\mathop{\smash{\operator@font E}\vphantom{a}}\nolimits}
\makeatother

\ifx\genfrac\sdflkaj

\else

\fi

\newcommand{\beq}{\begin{equation}}
\newcommand{\eeq}{\end{equation}}

\newtheorem{theorem}{Theorem}[section]
\newtheorem{lemma}[theorem]{Lemma}

\makeatletter
\def\mr@ignsp#1 {\ifx\:#1\@empty\else #1\expandafter\mr@ignsp\fi}%
\newcommand{\multiref}[1]{\begingroup%\let\protect\string%
\xdef\mr@no@sparg{\expandafter\mr@ignsp#1 \: }%
\def\mr@comma{}%
\@for\mr@refs:=\mr@no@sparg\do{\mr@comma\def\mr@comma{,}\ref{\mr@refs}}%
\endgroup}
\makeatother

\newcommand{\hypref}[2]{\ifx\href\asklfhas #2\else\href{#1}{#2}\fi}
\newcommand{\Secref}[1]{Section~\multiref{#1}}
\newcommand{\secref}[1]{Sec.~\multiref{#1}}

\newcommand{\tabref}[1]{Tab.~\multiref{#1}}

\newcommand{\figref}[1]{Fig.~\multiref{#1}}
\renewcommand{\eqref}[1]{(\multiref{#1})}

\def\[{\begin{equation}}
\def\]{\end{equation}}
\def\<{\begin{eqnarray}}
\def\>{\end{eqnarray}}

\ifx\href\asklfhas\newcommand{\href}[2]{#2}\fi

\title{On Dimensional Transmutation in 1+1D Quantum Hydrodynamics}

\author{Alexander Gorsky}
\address[Alexander Gorsky]{\newline
          Institute for Information Transmission Problems, \newline
          Russian Academy of Sciences, \newline
          Moscow, Russia \newline
          Email: \href{mailto:shuragor@mail.ru}{shuragor@mail.ru}
          \newline \href{http://iitp.ru/en/users/3804.htm}{http://iitp.ru/en/users/3804.htm}}

\author{Peter Koroteev}
\address[Peter Koroteev]{\newline
Department of Mathematics,\newline
University of California Berkeley,\newline
Evans Hall 970,\newline
Berkeley CA 94720,\newline
United States of America\newline
 Email: \href{mailto:pkoroteev@math.berkeley.edu}{pkoroteev@math.berkeley.edu}\newline 
 \href{https://math.berkeley.edu/~pkoroteev/}{https://math.berkeley.edu/~pkoroteev/}}

\author{Olesya Koroteeva}
\address[Olesya Koroteeva]{\newline
School of Physics and Astronomy,\newline
University of Minnesota\newline
Minneapolis MN 55455\newline
United States of America\newline
Email: \href{mailto:koroteeva@physics.umn.edu}{koroteeva@physics.umn.edu}\newline 
\href{https://www.physics.umn.edu/people/koroteeva.html}{https://www.physics.umn.edu/people/koroteeva.html}}

\author{Arkady Vainshtein}
\address[Arkady Vainshtein]{\newline
William Fine Theoretical Physics Institute and\newline
School of Physics and Astronomy,\newline
University of Minnesota.\newline
Minneapolis MN 55455,\newline
United States of America\newline
and Kavli Institute for Theoretical Physics,\newline
University of California Santa Barbara,\newline
Santa Barbara CA 93106, \newline
United States of America\newline
Email: \href{mailto:vainshte@umn.edu}{vainshte@umn.edu}\newline 
\href{https://www.physics.umn.edu/people/vainshte.html}{https://www.physics.umn.edu/people/vainshte.html}}

\date{\today}

\begin{document}

\begin{abstract}
Recently a detailed correspondence was established between, on one side, four and five-dimensional large-$N$  supersymmetric gauge theories with $\mathcal{N}\!=\!2$ supersymmetry and
adjoint matter, and, on the other side, integrable 1+1-dimensional quantum hydrodynamics. Under this correspondence the phenomenon of dimensional transmutation, familiar in asymptotically free QFTs, gets mapped to the 
transition from the elliptic Calogero-Moser many-body system to the closed Toda chain.
In this paper we attempt to formulate the hydrodynamical counterpart
of the dimensional transmutation phenomenon inspired by the identification of the 
periodic Intermediate Long Wave (ILW) equation as the hydrodynamical limit of the 
elliptic Calogero-Moser/Ruijsenaars-Schneider system. 
We also conjecture that the chiral flow in the vortex fluid provides the proper framework for the
microscopic description of such dimensional transmutation in the 1+1d hydrodynamics.
We provide a geometric description of this phenomenon in terms of the ADHM moduli space.
\end{abstract}

\maketitle
\setcounter{tocdepth}{1}
\tableofcontents

\section{Introduction and Main Results}
The integrable systems of Calogero-Ruijsenaars type emerged in the context of $\mathcal{N}=2$ supersymmetric 
gauge theories in four and five  dimensions with adjoint matter both at the classical 
\cite{Gorsky:1995zq,Martinec:1995by,Donagi:1995cf,nek97}
and the quantum levels \cite{Nekrasov:2009ui,Nekrasov:2009uh,Nekrasov:2009rc,Koroteev:2019gqi}. The quantum eigenvalue problem for these Hamiltonians can be formally solved by studying the dual gauge theory in the Omega-background in the Nekrasov-Shatashvili limit \cite{Nekrasov:2009rc}.
The degrees of freedom of the integrable system are provided by the defect surface operator added to the theory 
and its wave function corresponds to the instanton partition function in the presence of the defect.
In particular, the wave function of the elliptic Calogero-Moser model involves the Nekrasov equivariant integration
over the 4d instanton moduli space in the 4d $\mathcal{N}=2^*$ $U(N)$ gauge theory supplemented by the integral  over 2d instanton (vortex) moduli space in the theory on the defect. This is done by studying the moduli space of \textit{ramified} instantons \cite{Alday:2010vg,Nawata:2014nca,Bullimore:2015fr,Nekrasov:2017ab,pan,gfmn}) which uses equivariant integration over the affine 
Laumon space \cite{Negut:2011aa}. The coordinates of the maximal torus of the affine Laumon space are identified with the Fayet-Iliopoulos  couplings for the 2d defect theory.
The number of particles $N$ in the integrable system corresponds to the rank of the gauge group of the $\mathcal{N}=2^*$ theory 
while the spectral curve of the classical integrable system is nothing but the Seiberg-Witten curve of the dual gauge theory. 

It was shown some time ago \cite{2009JPhA...42m5201A } that in the large-$N$ limit the collective field theory can be developed for the trigonometric Calogero system which amounts to the hydrodynamical bidirectional Benjamin-Ono (BO) integrable system and, if the additional chiral constraint is imposed, 
the standard BO equation gets recovered. The generalization to the elliptic case has been found in \cite{Bonelli:2013rja,Bonelli:2014iza} (see \cite{Koroteev:2016,Koroteev:2018a} for the analysis in five dimensions and \cite{zabrodin} for the discrete version) using the relation with large-$N$ limit of the gauge theory.
It turned out that the hydrodynamical limit of the elliptic Calogero model is related to
the periodic ILW hydrodynamical equation which involves the elliptic kernel in the dispersion term. 
The BO equation can be derived from the ILW equation upon the trigonometric reduction of the elliptic kernel, while in a limit when the kernel reduces to the  $\delta(x-y)$ the KdV equation is recovered.

\subsection{Dimensional Transmutation and Many-Body Systems}

The supersymmetric $\mathcal{N}=2^*$ theory enjoys dimensional transmutation phenomenon whose 
origin is  the combination of the scale anomaly and the asymptotic freedom. The non-perturbative 
scale emerges at the quantum level in the classically conformal invariant theory 
\beq
\Lambda = M \exp\left(-\frac{4\pi}{\beta_0 g^2(M)}\right)\,,
\eeq
where $M$ is the regulator mass, which in our case corresponds to the mass of the adjoint hypermultiplet in the $\mathcal{N}=2^*$ theory,
and $\beta_0$ is the first coefficient in the expansion of the $\beta$-function in powers of the coupling $g$.
%\marginpar {\tiny How it's possible for regulator not to decouple?}The heavy regulator field does not decouple completely as $M\rightarrow \infty$ and provides the non-perturbative scale in the IR theory. 

The dimensional transmutation phenomenon has its counterpart in the world of the integrable many-body systems.
The elliptic Calogero-Moser model, whose potential exhibits long range interactions -- each particle interacts with every other particle: 
\begin{equation}
U_{\text{eCM}}(x_1,\dots x_n) \sim \sum_{i\neq j}\powerset(x_j-x_i)\,,
\label{eq:ElCalogero1}
\end{equation}
where $\powerset(x)$ is the Weierstrass function with elliptic parameter $\mathfrak{p}$ in the proper limit turns into the affine (periodic) Toda chain with potential
\begin{equation}
U_{\text{Toda}}(x_1,\dots x_n) \sim \sum_{i=1}^{N-1} e^{x_i-x_{i+1}} + \mathfrak{p}^\Lambda e^{x_N-x_1}\,,
\label{eq:Toda1}
\end{equation}
which has only near-neighbor interactions. This transformation is known as the Inozemtsev limit \cite{inozemtsev1989}
and involves the shift
$$x_i\to x_i + (i-1)\epsilon$$ 
when the system of particles sets a kind of 1d lattice and the Toda degrees of freedom 
correspond to the fluctuation of the lattice sites. (Note that in the gauge/integrability correspondence all variables are assumed to be complex).
Then one sends $\epsilon\to\infty$ leaving only the terms which differ by one unit of $\epsilon$
and scales the coupling constant. 

In this paper we shall question if there is  the hydrodynamical  analogue of the Inozemtsev limit thereby trying
to identify the phenomena of the dimensional transmutation in integrable hydrodynamics both at the 
classical and quantum levels. There are several ways to obtain the versions of Toda-like hydrodynamics
at the classical level. First, since we have only nearest-neighbor interaction in Toda, we could simply
rewrite the Hamiltonian equations of motion for Toda model using continuum variables. Along this way we get a version
of Toda hydrodynamics  which is analogous to bidirectional BO equations. In the other approach we could start with
the periodic ILW equation and consider an Inozemtsev-type limit 
of the elliptic kernel in the dispersion term of the equation itself.

The Inozemtsev limit can be discussed at the quantum level as well. In this case we can study limits of the integrals of motion of the quantum ILW equation directly. We shall describe a certain double scaling of the integrals of motion at the quantum level. Since at the gauge
theory side the hydrodynamics deals with the large-$N$ limit of the instanton counting we 
shall use this fact to describe the Inozemtsev limit in terms of the equivariant quantum cohomology/K-theory of the instanton moduli space.
In the elliptic case the instanton counting involves complicated interference
of 4d instantons and 2d instantons (vortices) on the defect. Instantons and vortices stay coupled
in the periodic Toda regime after the Inozemtsev limit is taken.
In limit $\mathfrak{p}\to 0$ the elliptic Calogero model \eqref{eq:ElCalogero1} 
is downgraded to its trigonometric version. 
On the gauge theory side this limit corresponds to perturbative regime -- the 4d instantons 
decouple leaving us with merely defect physics -- the affine Laumon space turns into the finite Laumon space.
The Inozemtsev limit of the trigonometric Calogero model yields non-periodic Toda potential 
\eqref{eq:Toda1} with $\mathfrak{p}^\Lambda$=0. 
We will reproduce the known statement that the wave function of the open Toda corresponds 
to the instanton counting in the 2d sigma-model that is 4d instantons get 
completely decoupled \cite{Gerasimov:835949}.

The origin of the relation between SYM instanton counting and the 1+1 hydrodynamics 
is the AGT correspondence.
According the AGT duality \cite{Alday:2009aq} and its generalization \cite{Wyllard:2009} 
the Nekrasov partition function
gets identified with the conformal blocks of the Liouville theory for the $SU(2)$
gauge theory and with $W_N$ blocks for the $SU(N)$ gauge theory. The insertion
of the surface operator at the gauge theory side amounts to the insertion 
of $\Psi_{2,1}$ in the conformal blocks where the number of insertions
corresponds to the number of Calogero particles. For the theory with adjoint matter
the conformal block on the torus has to be considered while in the pure
gauge theory the torus gets degenerated to a sphere. It is important 
that in terms of the $W_N$ algebra the asymptotically free gauge theory
without matter corresponds to the selection of the so-called Gaiotto
state or Whittaker vector which is the eigenvalue of $L_1$ -- the generator of the $SL(2,\mathbb{R})$
subalgebra \cite{Gaiotto:2009ma}. In the hydrodynamical limit we deals with $W_{1+\infty}$ 
that is our dimensional transmutation phenomena involves the analog 
of the Gaiotto state -- the coherent state in $SL(2,\mathbb{R})$ subalgebra of
$W_{1+\infty}$ or equivalently S$_\text{diff}$ -- algebra of area preserving diffeomorphisms.

\subsection{Towards dimensional transmutation in the vortex fluid}

Since dimensional transmutation occurs due to the combination 
of the scale anomaly and asymptotic freedom we have to search for both
in the hydro context. We suggest that the vortex fluid supporting the
chiral flow is the proper
candidate. There are two examples of the vortex fluid --- rotating
superfluid and the fractional quantum Hall effect (FQHE)  when
the bulk degrees of freedom are the interacting fermions at the lowest Landau level with an
attached flux (see i.e. \cite{Bernevig:2008aa,Bernevig:2008ac,Bernevig:2008ab}). Such composite particle presentation for the FQHE at the disc 
is consistent with the 
trigonometric Calogero model at the edge of the disc geometry \cite{PhysRevB.88.241305,Wiegmann2013}
and therefore to the bi-directional BO equation in the hydrodynamical limit.  
In general the BO equation is known to describe the edge 1+1 dynamics in the 2+1 vortex fluid.
It was argued recently that the boundary layer in the chiral
fluid has finite width and the vortex densities 
in the bulk and in the boundary layer are different \cite{Bogatskiy:2019}.
Remark that recently the relation between the Laughlin wave functions
on the torus and the instanton counting in the particular SYM gauge theory 
with defect has been discussed in \cite{Nekrasov_2019}.

The origin of this and some other non-trivial effects in the chiral flow in the vortex fluid
at the quantum level is the anomaly in the stress tensor found in \cite{Wiegmann:2019qvx}. 
The anomaly emerges if we take into account the UV cut-off
for the vortex size which is scale dependent.
Due to the scale anomaly vortices are no longer frozen in the flow since
the Helmgoltz law gets modified.
Hence we have in the vortex fluid the anomaly in the dilatation
transformation which is the necessary ingredient of the dimensional transmutation phenomena.

To recognize the second ingredient of the dimensional transmutation -- 
the asymptotic freedom for some running coupling constant we will look at   the interaction
between two vortices. Namely their finite sizes are taken into account via
the boundary conditions imposed on the wave function in two-body Calogero
system. Since the regulator mass in the gauge theory  plays the role of the coupling constant
for the vortex interaction in vortex fluid we get a kind of renormalization of the 
scale dependent UV cut-off. This can be presented in the form of 
the scale anomaly in the spectrum generating algebra in the 
Calogero model \cite{A_a_os_2003}.

Combining these arguments we will conjecture that the following picture emerges
which underlies the dimensional transmutation in hydro. At the
boundary between the bulk and the boundary layer of the vortex fluid due to
the conformal anomaly the vortices decouple from the fluid and form a 
kind of weakly fluctuating 1d lattice. On the other hand the `fermionic fluid'
(fermions in FQHE with fluxes detach) flow on the top of the vortex lattice 
interacting with phonons. This picture has a lot in common with the one
for the Peierls model of 1d superconductivity.

\subsection{Structure of the Paper}
In \Secref{Sec:Solitons1D} we shall review the classical ILW model and how its solitonic sector leads to the elliptic Calogero system. We shall briefly mention quantization and discuss scaling limit on the soliton ILW ansatz. The section ends with the review of the difference ILW model which is relevant for equivariant K-theory of the $U(1)$ instanton moduli space $\mathcal{M}_{k,1}$.
\Secref{Sec:ellRS} reviews the relationship between the eRS model and quantum Seiberg-Witten geometry of the $\mathcal{N}=1^*$ five-dimensional theory.
 In \Secref{Sec:InozemtsevRS} we describe the Inozemtsev limit of trigonometric and elliptic Ruijsenaars-Schneider models. 
This is followed by \Secref{Sec:quantumILW} which reminds the reader about the duality between quantum ILW and eRS models at large-$n$ along the lines of \cite{Koroteev:2018a, Koroteev:2018isw} and recollects the necessary information about equivariant K-theory of $\mathcal{M}_{k,1}$. In \Secref{Sec:InsttoVort} we study what happens with the Inozemtsev limit on the ILW side and formulate the new duality dictionary in which eRS model is replaced by the affine q-Toda chain.
In \Secref{Sec:PhysicsPicture} we shall address some questions concerning the microscopic aspects of the 
dimensional transmutation phenomena using the recently formulated quantum vortex fluid \cite{Wiegmann:2019qvx} as the starting point. Some open questions are formulated in \Secref{Sec:Conclusions}.

\subsection*{Acknowledgements}
We thank A. Abanov, A. Negut, and A. Zotov for the useful discussions and Kavli Institute for Theoretical Physics in Santa Barbara for support and hospitality.  
The research in KITP was supported in part by the National Science Foundation under Grant No. NSF PHY-1748958. 
The work of A.G. was also supported by Basis Foundation fellowship and RFBR grant 19-02-00214.
A.G. thanks Simons  Center for Geometry and  Physics, where this project was partly done, for support and hospitality. 
P.K. and O.K. thank Aspen Center for Physics, where this project was partly completed, for support and hospitality. 
P.K. is partly supported by AMS-Simons grant.

\section{Solitons in 1D ILW Hydrodynamics}\label{Sec:Solitons1D}
The ILW hydrodynamical model describes the spectrum of fluctuations on the interface of two fluid media inside a one-dimensional periodic channel. It can be represented by the following integro-differential equation
\begin{equation}
u_t + uu_{z}-\frac{\beta}{2}\, \partial_z^2 u^H =0\,,
\label{eq:ILWeq}
\end{equation}
where $u(x,t)$ is a complex-valued filed whose real and imaginary parts are related to the velocity and density fields of the fluid.  
Here the Hilbert-transformed field reads
\begin{equation}
u^H(z) = \frac{1}{2\pi}\text{v.p.}\,\int_0^{2\pi} \zeta(z'-z|\mathfrak{p}) u(z') dz'\,.
\end{equation}
With this normalization the periodic channel has length $2\pi$ and $\beta$ and $\mathfrak{p}$ are complex parameters which control the properties of the spectrum. In particular, $\mathfrak{p}$ is related to the ratio of the channel depth to the wavelength 
of perturbations as $\mathfrak{p}=e^{-2\pi\delta}$, where $\delta = h/\lambda$ is the ratio of the depth of the channel and the wavelength of fluctuations. In the shortwave limit $\mathfrak{p}\to 0$ the elliptic kernel becomes trigonometric thereby reducing the ILW equation into the Benjamin-Ono equation.

The ILW model is an integrable system with infinitely many integrals of motion
\begin{equation}
I_1 = \int_0^{2\pi} \frac{u^2}{2} dz \,,\quad I_2 = \int_0^{2\pi} \left[\frac{u^3}{3} +i\frac{\beta}{3}uu^H_z\right]dz\,,\dots\,,
\label{eq:ILWHamcl}
\end{equation}
so that the ILW equation \eqref{eq:ILWeq} can be written as Liouville evolution equation
\begin{equation}
u_t = \{u,I_2\}\,,
\end{equation}
with Hamiltonian $I_2$ and the integrals of motion are in involution $\{I_k,I_l\}=0$ with respect to canonical Poisson brackets.

\subsection{Pole Ansatz}
In this paper we focus on the dynamics of ILW solitons whose centers are located at coordinates $x_j(t)$. The total number of solitons is a topological invariant. In the sector with $k$ solitons the following pole Ansatz provides a solution of the ILW model. 
Let $u(z,t)=u_0(z,t)+u_1(z,t)$ where
\begin{equation}
\label{eq:defu0u1}
u_0(z,t) = i\beta\sum_{j=1}^k\widetilde\zeta(z-y_j(t)|\mathfrak{p})\,,\qquad u_1(z,t) = -i\beta\sum_{i=1}^k\widetilde\zeta(z-x_j(t)|\mathfrak{p})\,,
\end{equation}
where functions $x_j(t)$ and $y_j(t)$ satisfy
\begin{align}
\dot{x}_j&=u_0(x_j,t)+u^!_1(x_j,t)\,,\notag\\
\dot{y}_j&=u^!_0(y_j,t)+u_1(y_j,t)\,,
\label{eq:PoleAnsatzu}
\end{align}
where shrieks in the superscripts above designate the absence of terms with $i=j$ in the corresponding sums in \eqref{eq:defu0u1}, and where
\begin{equation}
\widetilde\zeta(\xi|\mathfrak{p})=\zeta(\xi|\mathfrak{p}) - \frac{2}{\pi}\eta_1\xi = \frac{\pi}{2\omega_1}\cot\left(\frac{\pi\xi}{2\omega_1}\right)+\frac{2\pi}{\omega_1}\sum_{l=1}^{\infty}\frac{\mathfrak{p}^{2l}}{1-\mathfrak{p}^{2l}}\sin\left(\frac{l\pi\xi}{\omega_1}\right)\,,
\label{eq:EllipticKernel}
\end{equation}
i.e., $\widetilde\zeta$ is the standard $\zeta$ function without the linear term. Note that if we have included the linear term we would have had $\delta^{-1}u_z$ term in the ILW equation. Note that $\widetilde\zeta = \frac{\theta_1'(\zeta|\mathfrak{p})}{\theta_1(\zeta|\mathfrak{p})}$, which was used, say in \cite{Bonelli:2014iza}. Now, if we denote $\widetilde u = u_0-u_1$ then the following equation holds
\begin{equation}
u_t + u u_z + \frac{i}{2}\beta\widetilde{u}_{zz}=0\,,
\end{equation}
which is equivalent to \eqref{eq:ILWeq} provided that $x_j$'s satisfy equations of motion for the \textit{elliptic Calogero-Moser-Sutherland} model for $k$ particles
\begin{equation}
\ddot{x}_j = -\beta^2\partial_j\sum_{i\neq j}\powerset(x_j-x_i)\,,\qquad i=1,\dots,k\,,
\label{eq:ElCalogero}
\end{equation}
where the Weierstrass $\powerset$ and $\zeta$ functions are related to each other via $\powerset(\xi)=-\frac{\partial}{\partial\xi}\zeta(\xi)$. Notice that the potential for the integrable many-body system is represented by the same function as in the pole ansatz for particles $x_j$ and momenta $y_j$\eqref{eq:defu0u1}.

\subsection{Quantization}
The model is also quantum integrable, this was studied in details earlier, see \cite{Koroteev:2016} and references therein. Complex velocity field $u$ can be expanded intro infinitely many oscillator modes $u(z,0)=\sum a_m e^{imz}$ which obey canonical commutation relations. 
The quantum ILW Hamiltonians which provide quantization of \eqref{eq:ILWHamcl} have the following form (see \cite{Koroteev:2018a} for review)
\begin{align} 
\widehat{I}_2 &= \sum_{m>0} a_{-m}a_m\,, \notag\\
\widehat{I}_3 &= \dfrac{\epsilon + m}{2} \sum_{m>0} m \dfrac{1 + (-\widetilde{\mathfrak{p}})^m}{1 - (-\tilde{\mathfrak{p}})^m} \, a_{-m} a_m + \dfrac{1}{2} \sum_{m,n > 0} (a_{-m-n} a_m a_n + a_{-m} a_{-n} a_{m+n})\,, 
\label{eq:I2I3ILW}
\end{align}
where $\epsilon=\log q$, $m=\log \hbar$, and $\widetilde{\mathfrak{p}}$ is the elliptic parameter.
The operators $a_{n}$ for negative $n$ create ILW solitons from the Fock vacuum $|0\rangle$ which is annihilated by all positive modes $a_{>0}|0\rangle = 0$. The operators $a_n$ obey the following commutation relations of the doubly-deformed Heisenberg algebra
\begin{equation}
[a_n,a_m]=m\frac{1-q^{m}}{1-\hbar^{m}}\delta_{m,-n}\,,
\label{eq:qHeisenb}
\end{equation}
where the deformation is a rational function of parameters $q$ and $\hbar$. In the semi-classical regime of the ILW model, when these two variables are expanded around unity this rational function becomes equal to $\epsilon/m$, which plays the role of the Planck's constant.

One can see how the scaling limit $\hbar\to\infty$ is manifest in the ILW pole Ansatz construction \eqref{eq:defu0u1} and \eqref{eq:PoleAnsatzu}.
Due to \eqref{eq:qHeisenb} we are required to rescale generators $a_n\to a_n \hbar^{-\frac{n}{2}}$ in this limit. If we return back to the oscillator representation of the velocity field $u$ we see that this rescaling entails shift in $z$-variable: $z\to z-i\frac{\epsilon}{2}$, where $\hbar=e^\epsilon$, in order to keep the decomposition $u(z,0)=\sum a_m e^{imz}$ in place. Additionally we put $\beta=\hbar\nu$, where $\nu$ is a nonzero constant which can be fixed later after we shall complete the quantum ILW computation. We shall return to analyzing the Inozemtsev limit of the free boson construction for the ILW model in \secref{Sec:GenFuncLambdaILW}.

\subsection{Modified Pole Ansatz}
At the classical level this procedure can be implemented in the pole Ansatz. 
First we need to modify the pole Ansatz configuration for $u$ into \eqref{eq:PoleAnsatzu}  as follows
\begin{equation}
\label{eq:defu0u1Mod}
u_0(z,t) = i\beta\sum_{j=1}^k\beta^{j-1}\widetilde\zeta(z-y_j(t)|\mathfrak{p})\,,\qquad u_1(z,t) = -i\beta\sum_{i=1}^k\beta^{j-1}\widetilde\zeta(z-x_j(t)|\mathfrak{p})\,,
\end{equation}
where we will assume that $|\beta|$ (and therefore $|\hbar|$) is large. Then we scale both $x_i$ and $y_j$ variables as
\begin{equation}
x_i\to x_i + \epsilon(i-1)\,,\qquad y_j\to y_j + \epsilon(j-1)\,,
\end{equation}
and by sending $\epsilon\to\infty$ we get 
\begin{align}
\dot{x}_j&=i\left[\sum_{i=1}^k e^{x_j-y_i}+\Lambda \delta_{j,1}e^{y_k-x_j}\right]-i \left[\sum_{i\neq j}^k e^{x_j-x_i}+\Lambda \delta_{j,1}e^{x_k-x_j}\right]\,,\notag\\
\dot{y}_j&=i\left[\sum_{i\neq j}^k e^{y_i-y_j}+\Lambda \delta_{j,1}e^{y_k-y_j}\right]-i \left[\sum_{i=1}^k e^{y_j-x_i(t)}+\Lambda \delta_{j,1}e^{x_k-y_j}\right]\,,
\label{eq:PoleAnsatzu2}
\end{align}
where $\Lambda = \mathfrak{p}\beta$ as $\beta\to \infty$ and $\mathfrak{p}\to 0$.
Notice that in this limit only near-neighbor interactions survive. 

The equations \eqref{eq:PoleAnsatzu2} are equivalent to the affine Toda equations of motion
\begin{equation}
\ddot{x}_j = -\partial_j \left[\sum\limits_{i=2}^{N}e^{x_i-x_{i-1}}+\Lambda e^{x_1-x_N}\right]\,.
\end{equation}

In the later sections we shall explore in great details how the above limit is manifest for quantum difference ILW model.

\subsection{Other Ways to Toda}
Let us discuss a different way to get to the Toda-like classical hydrodynamics.
First, we can perform the formal limiting procedure in ILW which yields the hyperbolic 
function from the $\zeta$-function in the kernel. We remind the reader
about the reduction of the Lame potential to the Matheu potential which
is an example of Inozemtsev limit for the two-body problem
\beq
M^2\powerset (x|\mathfrak{p})\rightarrow \Lambda^2 \cosh x
\eeq
The mass of the adjoint hypermultiplet in the gauge theory corresponds to the parameter
$\beta$ in the ILW model, hence the similar limiting procedure yields
\beq
\beta \zeta (x|\mathfrak{p}) \rightarrow \tilde{\Lambda} \sinh x 
\eeq
This procedure is quite formal and does not involve any microscopic degrees
of freedom at all. Nevertheless it yields the limit of the ILW equation obtained via the Inozemtsev-like limit. 
Strictly speaking this theory  probably has a different quantum description than 
the one we describe in the later sections, however it is worth exploring on its own.

Another microscopic derivation of the continuum limit  goes
as follows. Consider the finite Toda chain 
and introduce functions
\begin{equation}
x_n\rightarrow x(z)\,, \qquad p_n \rightarrow p(z)\,, \qquad x_{n-1} -x_n \propto \partial_z x(z)\,.
\end{equation}
The equations of motion can be presented in the form of the system 
of equations for the two functions \cite{297f7190af8743a989ed539c5a343b4a}
\begin{align}
2\partial_t \alpha &= (\alpha -\beta) \partial_z \alpha\,,\notag \\
2\partial_t \beta &=- (\alpha -\beta) \partial_z \beta\,,
\label{system}
\end{align}
where 
\beq
\alpha= p(z) + \partial_z x(z)\,, \qquad \beta= p(z) - \partial_z x(z)
\eeq
The periodicity in $z$ can be imposed by hands. 
Note that for $\beta=const$ we get the simple deformation
of the Hopf equation.

We have to check the Poisson
structure in the continuum limit. In the discrete case
we have evident Poisson bracket
\beq
\{p_i,x_j\}=\delta_{ij}
\eeq
The Poisson structure in the discrete case is properly inherited
in the continuum leading to the Kac-Moody symplectic structure
\beq
\{\alpha(z),\alpha(z')\}=\delta'(z-z')
\eeq
which is the correct Poisson structure for the Hopf equation.

The system (\ref{system}) is the Toda analogue 
of the bidirectional BO for Calogero model in the continuum.
In the Calogero case the BO equation can
be obtained from the bidirectional BO equation upon the
chiral reduction which selects only left or right movers.
Similar procedure has to be imposed for the (\ref{system}) 
if one would like to get the Toda limit of ILW equation.
We expect that the two approaches -- effective and microscopic  discussed in this subsection are 
related to each other. Hence one should be able to choose an appropriate chiral 
constraint for the bidirectional hydrodynamics such that the latter approach 
will be consistent with the former. We hope to study this issue in the near future.

\subsection{Difference ILW equation}
There is a difference version of the hydrodynamics -- $\Delta$ILW \cite{2009JPhA...42N4018S,zabrodin}, which is appropriate for the equivariant K-theory calculations. The kernel of the $\Delta$ILW integro-differential equation involves a finite-difference operator and reads
\begin{equation}
\dfrac{\partial}{\partial t} \eta (z,t) = \dfrac{i}{2} \eta (z,t)\, \text{v.p.} \int_{-1/2}^{1/2} 
(\Delta_{\gamma} \zeta)(\pi (w-z)) \cdot \eta(w,t) dw\,,
\label{fde}
\end{equation}
where the discrete Laplacian $\Delta_{\gamma}$ is defined as $(\Delta_{\gamma}f)(x)\! =\! f(x \!+\! \gamma)\! -\! 2f(x) \!+\! f(x\! -\! \gamma)$ and $\gamma$ is a complex number which is related to the radius of the compact circle of the dual gauge theory. In the limit $\gamma \rightarrow 0$ \eqref{fde} reduces to \eqref{eq:ILWeq}, after an appropriate Galilean transformation on field $\eta$. In \cite{Koroteev:2018a} it was shown, using the elliptic deformation of the Ding-Iohara algebra, that the quantum $\Delta$ILW system can be understood as the large-$n$ limit of quantum elliptic Ruijsenaars-Schneider model. In the next section we shall review this correspondence.

\subsection{ILW$_N$}
There is a non-Abelian generalization of the ILW system, as well as to its difference version which is referred to as ILW$_N$. It represents a fluid with non-Abelian velocity fields $u^{a}(z,t),\, a=1,\dots, N$ which interact with each other in a way that respects the $U(N)$ invariance (see \cite{lebedev1983,ref1,2013JHEP...11..155L,2015JHEP...02..150A} and \cite{2005PhRvL..95g6402A,2009JPhA...42m5201A} for the Benjamin-Ono limit). 

In \cite{Koroteev:2016}, using the connection with supersymmetric gauge theories, a relationship between the spectrum of $\Delta$ILW$_N$ and the moduli space of $U(N)$ instantons was established along the lines of the Abelian duality which we reviewed above. Not unexpectedly, the $\Delta$ILW$_N$ arises as a certain $n\to\infty$ limit of the 5d $U(Nn)$ $\CN=1^*$ gauge theory thereby providing a direct mapping between the parameters of both systems. Geometrically the $\Delta$ILW$_N$ Hamiltonians describe quantum multiplication in equivariant K-theory of $\mathcal{M}_{k,N}$.

\section{The Elliptic Ruijsenaars-Schneider Model}\label{Sec:ellRS}
In this section we review the formal solution of the elliptic RS model using quantum Seiberg-Witten geometry of the $\mathcal{N}=1^*$ 5d $U(n)$ theory with monodromy defect developed in \cite{Bullimore:2015fr}. Then we shall discuss in details the scaling (Inozemtsev) limit from eRS to closed qToda. We begin with the trigonometric RS model which describes physics on the 3d defect theory as well as its geometric meaning  \cite{Koroteev:2018isw}.

\subsection{Macdonald Difference Operators}
The difference operators of trigonometric Ruijsenaars-Schneider model with $n$ particles $\zeta_1,\dots,\zeta_n$ are given by
\begin{equation}
T_r(\vec{\zeta})=\sum_{\substack{\mathcal{I}\subset\{1,\dots,n\} \\ |\mathcal{I}|=r}}\prod_{\substack{i\in\mathcal{I} \\ j\notin\mathcal{I}}}\frac{\hbar^{-1/2}\,\zeta_i-\hbar^{1/2}\zeta_j}{\zeta_i-\zeta_j}\prod\limits_{i\in\mathcal{I}}p_k \,,
\label{eq:tRSRelationsEl}
\end{equation}
where $\vec{\zeta}=\{\zeta_1,\dots, \zeta_n\}$, the shift operator $p_k f(\zeta_k)=f(q\zeta_k)$.

It was proven in \cite{Koroteev:2018isw} that the vertex functions of the equivariant K-theory of the cotangent bundle to the complete flag variety, after proper normalization, is the eigenfunction of the tRS difference operators
\begin{equation}
\mathsf{V}_{\textbf{p}}= \prod\limits_{i=1}^n\frac{\theta(\hbar^{n-i}\zeta_i,q)}{\theta(a_i\zeta_i,q)}\cdot V^{(1)}_{\textbf{p}}\,,
\label{eq:DefVvertMac}
\end{equation}
where  
$$
\theta(x,q)=(x,q)_\infty (qx^{-1},q)_\infty = \prod\limits_{l=0}^\infty(1-q^l x) \prod\limits_{l=0}^\infty\left(1-\frac{q^{l+1}}{x}\right)
$$
is basic theta-function, while the vertex functions, which are labelled by the fixed points $\textbf{p}$ of the action of the maximal torus of $GL(n;\mathbb{C})$ 
\begin{equation}
V^{(1)}_{\textbf{p}}(z) = \sum\limits_{d_{i,j}\in C} \prod_{i=1}^{n-1} \left(t\frac{\zeta_{i}}{\zeta_{i+1}}\right)^{d_i} \prod\limits_{j,k=1}^{i}\frac{\left(q\frac{x_{i,j}}{x_{i,k}},q\right)_{d_{i,j}-d_{i,k}}}{\left(\hbar\frac{x_{i,j}}{x_{i,k}},q\right)_{d_{i,j}-d_{i,k}}}\cdot\prod_{j=1}^{i}\prod_{k=1}^{i+1}\frac{\left(\hbar\frac{x_{i+1,k}}{x_{i,j}},q\right)_{d_{i,j}-d_{i+1,k}}}{\left(q\frac{x_{i+1,k}}{x_{i,j}},q\right)_{d_{i,j}-d_{i+1,k}}}\,,
\label{eq:V1pdef}
\end{equation}
where $x_{n,k}=a_k$ and $t=\frac{q}{\hbar}$, $i=d_{i,1}+\dots+d_{i,i}$ and chamber C is determined via stability conditions of the quasimap. In other words, for each $i=1,\dots,n-2$  there should exist a subset in $\{d_{i+1,1},\dots d_{i+1,i+1}\}$ of cardinality $i$such that $d_{i,k} \geq d_{i+1,j_k}$. In the above formulae 
$$
(x,q)_d = \frac{(x,q)_\infty}{(q^d x,q)\infty}\,,\qquad (x,q)_\infty= \prod\limits_{l=0}^\infty(1-q^l x) \,.
$$

Vertex functions $V^{(\tau)}$ can be regarded as classes in equivariant K-theory of the moduli space of quasimaps 
\begin{equation}
\mathcal{H}_n:=K_{\textsf{T}}(\textbf{QM}(\mathbb{P}^1,X_n))
\label{eq:Kthquasimaps}
\end{equation}
for extended maximal torus $\textsf{T}$.

Then $\mathsf{V}_{\textbf{p}}$ are eigenfunctions for tRS difference operators \eqref{eq:tRSRelationsEl} for all fixed points $\textbf{p}$
\begin{equation}
T_r(\vec{\zeta}) \mathsf{V}_{\textbf{p}} = e_r (\mathbf{a}) \mathsf{V}_{\textbf{p}}\,, \qquad r=1,\dots, n\,,
\label{eq:tRSEigenz}
\end{equation}
where $e_r$ is elementary symmetric polynomial of degree $r$ of $a_1,\dots, a_n$\,.

It was then shown by one of the authors in \cite{Koroteev:2018isw} that vertex functions \eqref{eq:V1pdef} at the special locus 
\begin{equation}
\frac{a_{i+1}}{a_{i}}=q^{\ell_{i}}\hbar\,,\quad \ell_i = \lambda_{i+1}-\lambda_{i}\,, \quad i=1,\dots,n-1\,.
\label{eq:truncationa}
\end{equation}
truncate into symmetric Macdonald polynomials of $n$ variables $\vec{\zeta}$
\begin{equation}
\mathsf{V}_{\textbf{q}} = P_\lambda(\vec{\zeta};q,\hbar)\,,
\label{eq:MacPolyV}
\end{equation}
where $\lambda$ be a partition of $k$ elements of length $n$ and $\lambda_1\geq\dots\geq\lambda_n$.

\subsection{Elliptic Ruijsenaars-Schneider Model}
The Hamiltonians of the elliptic RS model can be easily obtained from trigonometric RS Hamiltonians \eqref{eq:tRSRelationsEl} by replacing rational functions with elliptic theta-functions of the first kind
\begin{equation}
E_r(\vec{\zeta})=\sum_{\substack{\mathcal{I}\subset\{1,\dots,n\} \\ |\mathcal{I}|=r}}\prod_{\substack{i\in\mathcal{I} \\ j\notin\mathcal{I}}}\frac{\theta_1(\hbar\zeta_i/\zeta_j|\mathfrak{p})}{\theta_1(\zeta_i/\zeta_j|\mathfrak{p})}\prod\limits_{i\in\mathcal{I}}p_k \,,
\label{eq:eRSRelationsEl}
\end{equation}
where $\mathfrak{p}\in\mathbb{C}^\times$ is the new parameter which characterizes the elliptic deformation away from the trigonometric locus, where $\mathfrak{p}=0$ and we get the trigonometric RS model Hamiltonians.

As in the trigonometric case we shall be interested in the eigenvalues and eigenfunctions of these operators
\begin{equation}
E_r(\vec{\zeta}) \mathcal{Z} = \mathscr{E}_r\mathcal{Z}\,,\qquad r=1,\dots,n\,.
\label{eq:eRSequation}
\end{equation}

It was verified in \cite{Bullimore:2015fr} that  the solution of \eqref{eq:eRSequation} is given by the K-theoretic holomorphic equivariant Euler characteristic of the affine Laumon space 
\begin{equation}
\mathcal{Z} = \sum_{\textbf{d}} \vec{\mathfrak{q}}^{\textbf{d}} \int\limits_{\mathcal{L}_{\textbf{d}}} 1\,,
\label{eq:equivaraintKthLaumon}
\end{equation}
where $\vec{\mathfrak{q}}=(\mathfrak{q}_1,\dots,\mathfrak{q}_n)$ is a string of $\mathbb{C}^\times$-valued coordinates on the maximal torus of $\mathcal{L}^{\text{aff}}_{\textbf{d}}$.
The eigenvalues $\mathscr{E}_r$ are equivariant Chern characters of bundles $\Lambda^r \mathscr{W}$, where $\mathscr{W}$ is the constant bundle of the corresponding ADHM space. In other words they have the following form
\begin{equation}
\mathscr{E}_r = e_r + \sum_{l=1}^\infty \mathfrak{p}^l \mathcal{E}^{(l)}_r \,.
\label{eq:eRSEnergies}
\end{equation}
The integrals in \eqref{eq:equivaraintKthLaumon} can be computed using localization and the resulting expression is an infinite sum over all sectors labelled by $k_l(\vec{\lambda})$
\begin{equation}
\mathcal{Z} = \sum_{\vec{\lambda}} \prod_{l=1}^n \mathfrak{q}_l^{k_l(\vec{\lambda})}\,z_{\vec{\lambda}}(\vec{a},\hbar, q)\,.
\label{eq:ZPartFuncRamFull}
\end{equation}
Here the topological sectors are defined as follows. The total number of boxes of $\vec{\lambda}=\{\lambda_{j,m}\}$ for $j=1,\dots, n$ and $m=1,\dots m_n$ adds up to $k=\sum _{l=1}^n k_l$ and $n = \sum_{m=1}^n m_n$.
The boxes in the $i$th column of Young diagram $\lambda_{j,k}$ contribute to the instanton sector $k_{i+j-1}$. If $i+j-1>n$ then we count modulo $n$ (see i.e. Sec. 4.2 of \cite{Bullimore:2015fr} for more details). 

In the limit when the parabolic structure is removed \eqref{eq:equivaraintKthLaumon} is expected to reproduced the well known Euler characteristic of $\mathcal{M}_N$ (Nekrasov instanton partition function) Thus we can impose the following
\begin{equation}
\mathfrak{p}=\mathfrak{q}_1\cdot\dots\cdot\mathfrak{q}_n\,,
\label{eq:ProdInst}
\end{equation}
where $\mathfrak{p}$ counts the degrees of sheaves in the standard ADHM localization computation.

The first several terms for the eigenvalues of $E_1$ look as follows
\begin{equation}
\mathscr{E}_1 = \sum_{i=1}^n a_i -\mathfrak{p}\hbar^{n}q^{-1}(1-\hbar^{-1})(q-\hbar^{-1})\sum_{i=1}^n a_i\prod_{\substack{j=1 \\ j\neq i}}^n \frac{(a_i-\hbar^{-1} a_j)(\hbar^{-1} a_i-q a_j)}{(a_i-a_j)(a_i-qa_j)}+o(\mathfrak{p}^2)\,.
\label{eq:E1eigneeRS}
\end{equation}

\section{Inozemtsev Limit in Ruijsenaars-Schneider Models}\label{Sec:InozemtsevRS}
Let us now discuss the scaling limit of the tRS and eRS models and their spectra.

\subsection{Quantum q-Toda System}
In \cite{Bullimore:2015fr} (Section 5.2) it was shown that the eigenfunction of $n$-body q-Toda Hamiltonians is given by a partition function $\mathcal{Z}^{\text{YM}}$ of pure $\mathcal{N}=1$ supersymmetric $U(n)$ Yang-Mills gauge theory on $\mathbb{C}_q\times\mathbb{C}\times S^1$ in the presence of the monodromy defect of maximal type wrapping $\mathbb{C}_q\times S^1$. 

This was established by studying limit $\hbar\to\infty$ in \eqref{eq:tRSEigenz} after certain rescaling also known as Inozemtsev limit \cite{inozemtsev1989}. First we rescale tRS coordinates, momenta \eqref{eq:tRSRelationsEl} and equivariant parameters $a_i$ as follows
\begin{equation}
\label{eq:InosemtsevScaling}
\mathfrak{z}_i= \hbar^{-i}\zeta_i\,,\qquad \mathfrak{p}_i = \hbar^{-i+1/2}p_i\,,\qquad \mathfrak{a}_i=\hbar^{-\frac{n}{2}}\alpha_i=a_i\,.
\end{equation}
After taking $\hbar\to\infty$ limit, we obtain q-Toda Hamiltonian functions which are equal to symmetric polynomials of $\mathfrak{a}_i$
\begin{equation}
H^{\text{q-Toda}}_r(\mathfrak{z}_1,\dots\mathfrak{z}_n;\mathfrak{p}_1,\dots, \mathfrak{p}_n)=e_r(\mathfrak{a}_1,\dots, \mathfrak{a}_n)\,,
\end{equation}
where the Hamiltonians are
\begin{equation}\label{eq:tRSRelationsElToda}
H^{\text{q-Toda}}_r=\sum_{\substack{\mathcal{I}=\{i_1<\dots<i_r\} \\ \mathcal{I}\subset\{1,\dots,n\}  }}\prod_{\ell=1}^r\left(1-\frac{\mathfrak{z}_{i_{\ell}-1}}{\mathfrak{z}_{i_\ell}}\right)^{1-\delta_{i_{\ell}-i_{\ell-1},1}}\prod\limits_{k\in\mathcal{I}}\mathfrak{p}_k \,,
\end{equation}
where $i_{0}=0$. For instance, the first Hamiltonian reads
\begin{equation}
H_1^{\text{open}}=\mathfrak{p}_1 +\sum\limits_{i=2}^n \mathfrak{p}_i \left(1-\frac{\mathfrak{z}_{i-1}}{\mathfrak{z}_{i}}\right)\,.
\end{equation}

\subsection{Inozemtsev Limit to Closed qToda}
For the elliptic RS model the Inozemtsev limit works as follows. The theta function has the following expansion near $\mathfrak{p}=0$
\begin{equation}
\theta_1(e^{iz}\vert\mathfrak{p})=2\mathfrak{p}^{\frac{1}{4}} \sum_{k=0}^{+\infty}(-1)^k \mathfrak{p}^{k (k+1)} \sin ((k+1/2) z)\,,
\end{equation}
The eRS Hamiltonians \eqref{eq:eRSRelationsEl} contain the following ratio of theta-functions which have the following expansion around $\mathfrak{p}=0$
\begin{equation}
\frac{\theta_1\left(\frac{\zeta_1}{\hbar\zeta_2}\vert\mathfrak{p}\right)}{\theta_1\left(\frac{\zeta_1}{\zeta_2}\vert\mathfrak{p}\right)}=
\frac{\frac{\sqrt{\frac{\zeta_1}{\zeta_2}}}{\sqrt{\hbar }}-\frac{\sqrt{\hbar}}{\sqrt{\frac{\zeta_1}{\zeta_2}}}+\mathfrak{p}^2 \left(\frac{\left(\frac{\zeta_1}{\zeta_2}\right){}^{3/2}}{\hbar ^{3/2}}-\frac{\hbar
   ^{3/2}}{\left(\frac{\zeta_1}{\zeta_2}\right){}^{3/2}}\right)}{\sqrt{\frac{\zeta_1}{\zeta_2}}-\frac{1}{\sqrt{\frac{\zeta_1}{\zeta_2}}}+\mathfrak{p}^2
   \left(\frac{1}{\left(\frac{\zeta_1}{\zeta_2}\right){}^{3/2}}-\left(\frac{\zeta_1}{\zeta_2}\right){}^{3/2}\right)}+O(\mathfrak{p}^5)
\end{equation}

After taking the limit $\hbar\to\infty$ the above formula after applying scaling \eqref{eq:InosemtsevScaling} the two-body eRS Hamiltonian reads
\begin{equation}
\frac{\theta_1\left(\frac{\zeta_1}{\hbar\zeta_2}\vert\mathfrak{p}\right)}{\theta_1\left(\frac{\zeta_1}{\zeta_2}\vert\mathfrak{p}\right)}p_1+\frac{\theta_1\left(\frac{\zeta_2}{\hbar\zeta_1}\vert\mathfrak{p}\right)}{\theta_1\left(\frac{\zeta_1}{\zeta_2}\vert\mathfrak{p}\right)}p_2\to 
\mathfrak{p}_1 \left(1-\mathfrak{q}\frac{\mathfrak{z}_2}{\mathfrak{z}_1}\right)+\mathfrak{p}_2 \left(1-\frac{\mathfrak{z}_1}{\mathfrak{z}_2}\right)\,,
\end{equation}
where we assumed $\mathfrak{q} = -\mathfrak{p}^2 \hbar^{2}$ is finite. The new term proportional to $\mathfrak{q}$ arises which ensures periodicity. For an $n$-body eRS model we get the following formula for the first affine q-Toda Hamiltonian
\begin{equation}
H^{\text{aff q-Toda}}_1=\mathfrak{p}_1\left(1-\mathfrak{q}\frac{\mathfrak{z}_{n}}{\mathfrak{z}_1}\right) +\sum\limits_{i=2}^n \mathfrak{p}_i \left(1-\frac{\mathfrak{z}_{i-1}}{\mathfrak{z}_{i}}\right)\,,
\end{equation}

\subsection{Spectrum of Closed qToda}
One gets the following equations for the spectrum of quantum closed q-Toda
\begin{equation}
H^{\text{aff q-Toda}}_r(\vec{\zeta}) \mathcal{Z}^{\text{YM}} = \mathscr{E}^{Toda}_r\mathcal{Z}^{\text{YM}}\,,\qquad r=1,\dots,n\,,
\label{eq:eRSequationToda}
\end{equation}
and $\mathscr{E}^{Toda}_r$ is given by the $\hbar\to\infty,\,\mathfrak{p}\to 0$ limit of the eRS energies $\mathscr{E}_r$
\begin{equation}
\mathscr{E}^{\text{Toda}}_1 = \sum_{i=1}^n \mathfrak{a}_i +\mathfrak{q}\sum_{i=1}^n \mathfrak{a}_i\prod_{\substack{j=1 \\ j\neq i}}^n \frac{1}{\left(1-\frac{a_j}{a_i}\right)\left(1-\frac{a_i}{qa_j}\right)}+O(\mathfrak{q}^2)\,,
\label{eq:E1eigneeRSToda}
\end{equation}
where $\mathfrak{q}=\mathfrak{p}\hbar^n$.

\section{Quantum $\Delta$ILW Spectrum}\label{Sec:quantumILW}
Let us first describe the Hilbert space of the quantum ILW. The cohomological version was studied in \cite{Okounkov:2004a}, here work in equivariant K-theory.

We can map Macdonald polynomials to states in the Fock space representation of the $q,\hbar$-Heisenberg algebra \eqref{eq:qHeisenb} by claiming that
\begin{equation}
x_k = a_{-k}\vert 0\rangle\,,
\label{eq:PowerSymBasis}
\end{equation}
where $x_k = \sum_{l=1}^n \zeta_l^k$. In this symmetric basis polynomials $P_\lambda$ only depend on the number of boxes of tableau $\lambda$ -- $k$ do not explicitly depend on $n$. Such Macdonald polynomials form a basis in the equivariant K-theory of Hilbert schemes of $k$ points on $\mathbb{C}^2$. See \cite{2011ntqi.conf..357S,2013arXiv1309.7094S,2013arXiv1301.4912S} for more details.

The identification \eqref{eq:PowerSymBasis} allows us to make the following matching  
\begin{theorem}[\cite{Koroteev:2018isw}]
\label{Th:EmbeddingTh}
For $n>k$ there is the following embedding of Hilbert spaces
\begin{align}
\bigoplus\limits_{l=0}^k K_{q,\hbar}(\text{Hilb}^l(\mathbb{C}^2)) &\hookrightarrow \mathcal{H}_n
\label{eq:Embeddingn}\\
[\lambda]&\mapsto \mathsf{V}_{\textbf{q}}\,.\notag
\end{align}
for K-theory vertex function for some fixed point $q$ of maximal torus $T$ evaluated at locus \eqref{eq:truncationa}.
The statement also holds in the limit $n\to\infty$
\begin{equation}
\bigoplus\limits_{l=0}^\infty K_{q,\hbar}(\text{Hilb}^l(\mathbb{C}^2)) \hookrightarrow \mathcal{H}_\infty\,,
\label{eq:Embeddinginf}
\end{equation}
where $\mathcal{H}_\infty$ is defined as a stable limit of $\mathcal{H}_n$ \eqref{eq:Kthquasimaps} as $n\to\infty$.
\end{theorem}

Now let us talk about the ILW spectrum. 

\subsection{Quantum Benjamin-Ono Spectrum}
In the Benjamin-Ono limit ($\tp \to 0$) the spectrum can be realized in terms of the tRS eigenvalues.
The eigenvalues of BO Hamiltonians can be realized geometrically as operator of multiplication by the universal bundle to the moduli space of $U(1)$ instantons.
\begin{lemma}[\cite{Koroteev:2018isw}]
The eigenvalue of the operator of multiplication by the universal bundle corresponding to $\mathscr{V}\vert_{\mathcal{J}_\lambda}$ 
\begin{equation}
\mathscr{U} = \mathscr{W}-(1-\hbar)(1-q)\mathscr{V}\vert_{\mathcal{J}_\lambda}\,,
\end{equation}
where $\mathscr{W}$ is a constant bundle of degree 1, in equivariant K-theory $K_{q,\hbar}(\text{Hilb}^k)$ is given by
\begin{equation}
\mathcal{E}_1(\lambda) = a\left(1-(1-\hbar)(1-q)\sum\limits_{(i,j)\in\lambda} \sum_{c=1}^k s_c\right)\,,
\label{eq:EUnivBundleEigen1}
\end{equation}
where $s_1,\dots s_k$ are in one-to-one correspondence with the content of tableau $\lambda$ of size $k$ corresponding to class $[\mathcal{J}]$ and are given by
\begin{equation}
s_{i,j} = q^{i-1}\hbar^{j-1}\,,
\end{equation}
for $i,j$ ranging through the co-arm and co-leg of $\lambda$. In \eqref{eq:EUnivBundleEigen1} $a\in\mathbb{C}^\times$ is the character of $T\mathscr{W}$. 
\end{lemma}

The spectrum of the ILW model will also include elliptic deformations, which we shall review below.

\subsection{Quantum $\Delta$ILW Spectrum}
The quantum difference ILW ($\Delta$ILW) Hamiltonian can be constructed by combining quantum ILW operators \eqref{eq:I2I3ILW} as follows \cite{Feigin:2009ab,Koroteev:2016,Koroteev:2018a}
\begin{equation}
\mathcal{H}_{\text{ILW}} = [\eta(\xi)]_1 = 1 +  \widehat{I}_2 + \widehat{I}_3 + \widehat{I}_4 + \ldots \label{gf}\,,
\end{equation}
where subscript $1$ shows that we need to pick a term in front of $\xi^1$ of the following generating function 
\begin{equation}
\eta(\xi) = \text{exp}\left(\sum_{n>0}\lambda_{-n}\xi^n\right) \text{exp}\left(\sum_{n<0}\lambda_{n}\xi^{-n}\right)\,, 
\label{eq:genfuncILW}
\end{equation}
in which the raising and lowering operators obey the following triply-deformed Heisenberg algebra (See, i.e. \cite{Koroteev:2018a}, Appendix B for details).
\begin{equation}
[\lambda_m, \lambda_n] = -\dfrac{1}{m} \dfrac{(1-q^{m})(1-\hbar^{m})(1-(-\tp q^{-1/2}\hbar^{-1/2})^{m})}{1-(-\tp q^{1/2}\hbar^{1/2})^{m}} \delta_{m+n,0}\,.
\label{eq:TripleHeis}
\end{equation}
The oscillators can be normalized as
\begin{equation}
\lambda_m = \dfrac{1}{\vert m \vert} \sqrt{-\dfrac{(1-q^{\vert m \vert})(1-\hbar^{\vert m \vert})(1-(-\tp q^{-1/2}\hbar^{-1/2})^{\vert m \vert})}{1-(-\tp q^{1/2}\hbar^{1/2})^{\vert m \vert}}} a_m\,,
\end{equation}
with commutation relations
\begin{equation}
[a_m, a_n] = m \delta_{m,-n}\,.
\end{equation}

The operator \eqref{eq:genfuncILW} can be used to describe the eRS Hamiltonians in the free boson formalism. For instance, the realization of the first Hamiltonian $E_1$ \eqref{eq:eRSRelationsEl} acting on the state 
\begin{equation}
\Psi_n (\vec{\zeta}) |0\rangle = \prod\limits_{a=1}^n \exp\left(\sum\limits_{j>0}\frac{1}{j}\frac{1-\hbar^j}{1-q^j}a_{-j}\zeta_a^j\right) |0\rangle
\end{equation}
of the Hilbert space is given by
\begin{equation}
[\eta(\xi;\tp)]_1\Psi_n (\vec{\zeta}) |0\rangle = \Psi_n (\vec{\zeta}) [P_n(\vec{\zeta})\eta(\xi;\tp)]_1 |0\rangle+\hbar^{n-1}(1-\hbar)S(\tp,q,\hbar) E_1(\zeta,\widetilde{\mathfrak{p}})|0\rangle\,,
\label{eq:EllRSFreeBoson}
\end{equation}
where
\begin{equation}
P_n(\vec{\zeta})=\hbar^{-n}\prod\limits_{i=1}^n\frac{\theta_1\left(q\hbar\frac{\xi}{\zeta_i}\Big\vert\widetilde{\mathfrak{p}}\right)}{\theta_1\left(q\frac{\xi}{\zeta_i}\Big\vert\widetilde{\mathfrak{p}}\right)}\frac{\theta_1\left(\hbar^{-1}\frac{\xi}{\zeta_i}\Big\vert\widetilde{\mathfrak{p}}\right)}{\theta_1\left(q\frac{\xi}{\zeta_i}\Big\vert\widetilde{\mathfrak{p}}\right)}\,,\qquad S(\tp,q,\hbar)=\frac{(\tp\hbar;\tp)_\infty (\tp\hbar q^{-1};\tp)_\infty}{(\tp;\tp)_\infty(\tp q^{-1};\tp)_\infty}\,.
\end{equation}
Due to the presence of the first term in the right hand side of \eqref{eq:EllRSFreeBoson} $\Psi_n (\vec{\zeta}) |0\rangle$ is not an eigenstate of $[\eta(\xi)]_1$, so there is no one-to-one correspondence between the eigenvalues of $[\eta(\xi)]_1$ and $E_1$ at finite $n$.
However, this term disappears in the $n\to 0$ limit provided that $|\hbar|>1$ which is exactly our situation since $|\hbar|$ is large in the Inozemtsev limit. This allows us to recover information about the spectrum of the $\Delta$ILW model starting from the eRS system with the large-$N$umber of particles. 

It was then demonstrated in \cite{Koroteev:2016,Koroteev:2018a} that the eigenvalue of $[\eta(\xi)]_1$in \eqref{eq:EllRSFreeBoson} is related to the operator of quantum multiplication by the universal bundle over the instanton moduli space $\mathcal{M}_{k,1}$
\begin{equation}
\mathscr{U} = \mathscr{W}-(1-\hbar)(1-q)\mathscr{V}
\end{equation}
in the quantum equivariant K-theory $K_{q,\hbar}(\text{Hilb}^k(\mathbb{C}^2))$. The corresponding eigenvalue is given by the following formula
\begin{equation}
\mathcal{E}_1(\lambda) = 1-(1-\hbar)(1-q) e_1(s_1,\dots,s_k)\,,
\label{eq:EUnivBundleEigen}
\end{equation}
where $e_1(s_1,\dots,s_k)=s_1+\dots+s_k$ is the 1st elementary symmetric polynomial of $s_1,\dots s_k$ which solve the following Bethe equations, which are applicable for the quantum spectrum of the $\Delta$ILW$_N$ model. ($N=1$ for the $\Delta$ILW)
\begin{equation}
\prod_{l=1}^N\frac{s_a-\mathrm{a}_l}{s_a-q\hbar\mathrm{a}_l}\cdot\prod_{\substack{b=1 \\ b\neq a}}^k\frac{s_a-q^{-1} s_b}{s_a-qs_b}\frac{s_a-\hbar^{-1} s_b}{s_a-\hbar s_b}\frac{s_a-q\hbar s_b}{s_a-(q\hbar)^{-1} s_b}=\widetilde{\mathfrak{p}}(-q^{1/2}\hbar^{1/2})^N\,,\quad a=1,\dots, k\,,
\label{eq:BetheADHM}
\end{equation}
where $\widetilde{\mathfrak{p}}$ is the K\"{a}hler parameter of the ADHM quiver and is related to the 5d instanton counting parameter $\mathfrak{p}$ as follows \cite{Koroteev:2016}
\begin{equation}
\widetilde{\mathfrak{p}}(-q^{1/2}\hbar^{1/2})^N=\mathfrak{p}q^N\hbar^N\,.
\end{equation}

\begin{figure}[!h]
\begin{tikzpicture}[scale =1.2]
\draw [ultra thick] (0,0) -- (3,0);
\draw [ultra thick] (3,1) -- (3,0);
\draw [fill] (0,0) circle [radius=0.1];
\draw [fill] (1,0) circle [radius=0.1];
\draw [fill] (2,0) circle [radius=0.1];
\draw [fill] (3,0) circle [radius=0.1];
\node (1) at (0.1,-0.3) {$\mathbf{1}$};
\node (2) at (1.1,-0.3) {$\mathbf{2}$};
\node (3) at (2.1,-0.3) {$\ldots$};
\node (4) at (3.1,-0.3) {$\mathbf{n-1}$};
\fill [ultra thick] (3-0.1,1) rectangle (3.1,1.2);
\node (5) at (3.1,1.45) {$\mathbf{n}$};
\end{tikzpicture}
\qquad \qquad \qquad
\begin{tikzpicture}[xscale=1.5, yscale=1.5]
\fill [ultra thick] (1-0.1,1) rectangle (1.1,1.2);
\draw [fill] (1,0) circle [radius=0.1];
\draw [-, ultra thick] (1,1) -- (1,0.1);
\draw[-, ultra thick]
(1,0) arc [start angle=90,end angle=450,radius=.3];
\node (1) at (1.4,1.15) {$\mathscr{W}$};
\node (2) at (1.4,0.2) {$\mathscr{V}$};
\end{tikzpicture}
\caption{\small Left: Quiver diagram for the cotangent bundle to the complete flag variety $X_n=T^*\mathbb{F}l_n$.  Right: The ADHM quiver. Undirected links between nodes depict 3d $\mathcal{N}=4$ hypermultiplets.}
\label{fig:QuiverVarieryNak}
\end{figure}
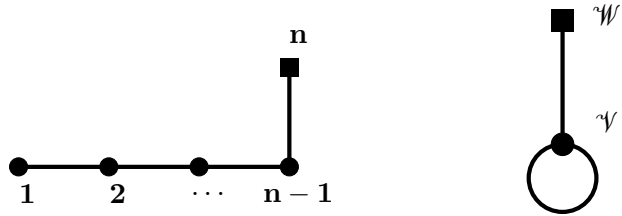

The equations \eqref{eq:BetheADHM} describe the Coulomb branch of the 3d $\mathcal{N}=2^*$ $U(k)$ theory with $N$ hypermultiplets (see the right picture in \figref{fig:QuiverVarieryNak}). In more details the matter content of the ADHM quiver gauge theory is summarized in the table below 
\begin{table}[h!]
\begin{center}
\begin{tabular}{|c|c|c|c|c|c|}
\hline
Fields & $\chi$ & $B_{1}$ & $B_{2}$ & $I$ & $J$ \\[1mm] \hline\hline
gauge group $U(k)$ & $\text{Adj}$ & $\text{Adj}$ & $\text{Adj}$ & $\mathbf{k}$ & $\mathbf{\bar{k}}$ \\[1mm] \hline
flavor $U(N)\times U(1)^{2}$ & $\mathbf{1}_{(-1,-1)}$ & $\mathbf{1}_{(1,0)}$ & $\mathbf{1}_{(0,1)}$ & $\mathbf{\bar{N}}_{(0,0)}$ & $\mathbf{N}_{(1,1)}$ \\[1mm] \hline
flavor parameters & $(q\hbar)^{-1}$ & $q$ & $\hbar$ & $\mathrm{a}_{j}$ & $\mathrm{a}_{j}^{-1}q\hbar$ \\[1mm] \hline
$R$-charge & $2$ & $0$ & $0$ & $0$ & $0$ \\[1mm] \hline
\end{tabular} 
\vspace{2mm}
\caption{\small Matter content of the ADHM 3d quiver theory.}
\label{tab:TableFields}
\end{center}
\end{table} 

Additionally there is a superpotential which is given by $W=\textrm{Tr}_{k}\left\{\chi\left([B_{1},B_{2}]+IJ\right)\right\}$. 
Notice that the product of the flavor fugacities of fields $\chi, B_1$ and $B_2$ is equal to one (equivalently, the sum of their twisted masses vanishes). This property arises from the Calabi-Yau compactification of the underlying string geometry \cite{Bonelli:2014iza}.

\subsection{The Gauge/Hydrodynamics Correspondence}
It was show in \cite{Koroteev:2016,Koroteev:2018a} that large-$n$ limit of the VEV of the Wilson loops in 5d $\mathcal{N}=1^*$ theory are proportional to characters of the universal bundle on the tangent bundle to the moduli space of $U(1)$ instantons evaluated on the locus \eqref{eq:aequivspec}, in particular, in case of the fundamental Wilson loop we get
\begin{equation}
\lim_{n\to\infty} \left[\hbar^{n-1}(1-\hbar)\left\langle W_{\tiny{\yng(1)}}^{U(n)}\right\rangle\right]\Big\vert_\lambda = a-(1-q)(1-\hbar) e_1(s_1,\dots,s_k) \vert_\lambda\,.
\label{eq:BPSHydroCorr}
\end{equation}
In other words, the ILW energies \eqref{eq:EUnivBundleEigen} evaluated at the solutions of Bethe equations \eqref{eq:BetheADHM} are equal to the eRS energies \eqref{eq:E1eigneeRS} on the locus
\begin{equation}
a_i = a q^{\lambda_i}\hbar^{i-n}\,,\quad i=1,\dots,n\,, 
\label{eq:aequivspec}
\end{equation}
where $|\lambda|=k$ in the limit when $n\to\infty$. The summary of the correspondence is given in \tabref{Tab:Corresp}.

\begin{table}[!h]
\begin{center}
\renewcommand\arraystretch{1.2}
\begin{tabular}{|c|c|c|}
\hline
Elliptic RS model& 5d/3d $\mathcal{N}=2^*$ theory & 3d ADHM theory  \\ 
\hline\hline Coordinates $z_i$  & K\"ahler parameters & K-ring generators $x_i$ \\ 
\hline Eigenstates $\lambda$ & Defect partition functions & ADHM Coulomb branch vacua \\
\hline Planck constant $\log q$& equivariant parameter $q$& $\mathbb{C}^\times_q$ acting on $\mathbb{C}\subset\mathbb{C}^2$  \\
\hline Coupling constant $\hbar$ &$\mathbb{C}^\times_\hbar$ acting on cotangent fibers of $X$ & $\mathbb{C}^\times_\hbar$ acting on another $\mathbb{C}\subset\mathbb{C}^2$  \\
\hline Elliptic parameter $\mathfrak{p}$ & 5d gauge coupling $e^{-\frac{8\pi^2}{g_{\text{YM}}^2}}$  & FI coupling $-\mathfrak{p}/\sqrt{q\hbar}$ \\
\hline  Eigenvalues $\mathscr{E}_r$ & VEVs of Wilson loop $\langle W_{\Lambda^r \tiny{\yng(1)}}\rangle$ & Chern polynomials $\mathcal{E}_r$ of $\Lambda^r\mathcal{U}$  \\
\hline
\end{tabular}
\vspace{2mm}
\renewcommand\arraystretch{1}
\caption{\small The correspondence table between the elliptic RS model, its 5d/3d gauge theory description and large-$n$ ADHM quiver description.}
\label{Tab:Corresp}
\end{center}
\end{table}

For the correspondence which involves $\Delta$ILW$_N$ model we start with $U(nN)$ $\mathcal{N}=2^*$ gauge theory, proceed similar to the above and replace the locus \eqref{eq:aequivspec} with
\begin{equation}
a_\alpha = \widetilde{\mathrm{a}_\alpha}\, q^{\Lambda_\alpha}\, \hbar^{\alpha-nN}\,,\qquad \alpha = 1,\dots, nN\,,
\end{equation}
where  
\begin{equation}
\{\widetilde{\mathrm{a}_\alpha}\} = \{\mathrm{a}_1,\dots \mathrm{a}_1,\mathrm{a}_2,\dots,\mathrm{a}_2,\dots,\mathrm{a}_N,\dots,\mathrm{a}_N\}
\end{equation}
and Young tableau $\Lambda$ is blended from $N$ diagrams $\Lambda=\lambda_1\circledast\ldots\circledast\lambda_N$. We refer the interested reader to Sec. 4.1 of \cite{Koroteev:2018isw} for details.

\section{From Instantons to Vortices}\label{Sec:InsttoVort}
Now we shall describe how the Inozemtsev limit is implemented on the resolved side of the duality. In summary, under the $\hbar\to\infty$ limit the instanton moduli space $\mathcal{M}^{\text{inst}}_{1,k}$ will get retracted to the vortex moduli space $\mathcal{M}^{\text{vort}}_{1,k}$.

\subsection{Scaling Limit}
\begin{figure}[!h]
\begin{tikzpicture}[scale =1.2]
\draw [thick] (0,0) -- (3,0);
\draw [->, thick] (0,0) -- (0.5,0);
\draw [->, thick] (1,0) -- (1.5,0);
\draw [->, thick] (2,0) -- (2.5,0);
\draw [thick] (3,1) -- (3,0);
\draw [->, thick] (3,0) -- (3,0.6);
\draw [fill] (0,0) circle [radius=0.1];
\draw [fill] (1,0) circle [radius=0.1];
\draw [fill] (2,0) circle [radius=0.1];
\draw [fill] (3,0) circle [radius=0.1];
\node (1) at (0.1,-0.3) {$\mathbf{1}$};
\node (2) at (1.1,-0.3) {$\mathbf{2}$};
\node (3) at (2.1,-0.3) {$\ldots$};
\node (4) at (3.1,-0.3) {$\mathbf{n-1}$};
\fill [ultra thick] (3-0.1,1) rectangle (3.1,1.2);
\node (5) at (3.1,1.45) {$\mathbf{n}$};
\end{tikzpicture}
\qquad \qquad \qquad
\begin{tikzpicture}[xscale=1.5, yscale=1.5]
\fill [ultra thick] (1-0.1,1) rectangle (1.1,1.2);
\draw [fill] (1,0) circle [radius=0.1];
\draw [thick] (1,1) -- (1,0.5) ;
\draw [->, thick] (1,0.1) -- (1,0.5);
\draw[->, thick]
(1,0) arc [start angle=90,end angle=270,radius=.3];
\draw[thick]
(1,0) arc [start angle=90,end angle=450,radius=.3];
\node (1) at (1.4,1.15) {$\mathscr{W}$};
\node (2) at (1.4,0.2) {$\mathscr{V}$};
\end{tikzpicture}
\caption{\small Left: Quiver diagram for the complete flag variety $\mathbb{F}l_n$. Right: The $\frac{1}{2}$-ADHM quiver. Chiral multiplets are depicted with arrows.}
\label{fig:QuiverVarieryNak1}
\end{figure}
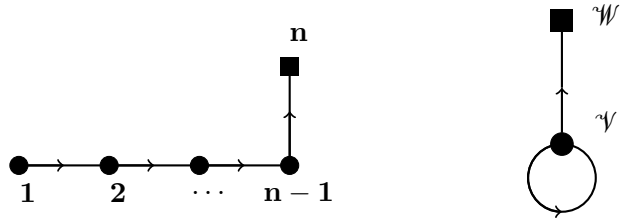

We can take $\hbar\to\infty$ limit of the above formulae in the ADHM construction. We get the following for $r=1$
\begin{equation}
\mathcal{E}^{\Lambda}_1(\lambda) = a-(1-q) e_1(s_1,\dots,s_k)\,,
\label{eq:EUnivBundleEigennew}
\end{equation}
where Bethe roots now solve the equations arising from the vortex moduli space ($N$ chirals and one chiral loop)
\begin{equation}
\prod_{l=1}^N(s_a-\mathrm{a}_l)\cdot\prod_{\substack{b=1 \\ b\neq a}}^k\frac{q s_a- s_b}{s_a-q s_b}=\widetilde{\mathfrak{p}}^{\Lambda}\,,\quad a=1,\dots, k\,,
\label{eq:BetheADHMnew}
\end{equation}
where 
\begin{equation}
\widetilde{\mathfrak{p}}^{\Lambda}=\widetilde{\mathfrak{p}}\,q^{1/2}\hbar^{1/2}\prod_{l=1}^N(-q\hbar a_l)
\label{eq:pLambdadef}
\end{equation}
is the dynamically generated scale and the quantum parameter. The above equations \eqref{eq:BetheADHMnew} describe the Coulomb branch of the 3d $\mathcal{N}=2$ $U(k)$ theory with $N$ chiral multiplets (see right figure in \figref{fig:QuiverVarieryNak1}).

\subsection{The Gauge/Hydrodynamics Correspondence in the $\hbar\to\infty$ limit}
Having taken the $\hbar\to\infty$ limit on both sides of the correspondence \eqref{eq:E1eigneeRSToda} and \eqref{eq:EUnivBundleEigennew} we arrive to our main conclusion
\begin{equation}
\lim_{n\to\infty} \hbar^{n}\mathscr{E}^{\text{Toda}}_1\Big\vert_\lambda = \mathcal{E}^{\Lambda}_1(\lambda) \vert_\lambda\,,
\label{eq:BPSHydroCorr2}
\end{equation}
where the equivariant parameters on the left hand side for the q-Toda eigenvalues \eqref{eq:E1eigneeRSToda} on the locus 
\begin{equation}
\mathfrak{a}_i = a q^{\lambda_i}\,,\quad i=1,\dots,n\,, 
\label{eq:aequivspectoda}
\end{equation}
while the $\Delta$ILW energies \eqref{eq:EUnivBundleEigennew} are evaluated on the solutions of scaled Bethe equations \eqref{eq:BetheADHMnew}. The instanton counting parameters from \eqref{eq:E1eigneeRSToda} and  \eqref{eq:pLambdadef} are then identified as
\begin{equation}
\mathfrak{q}=\widetilde{\mathfrak{p}}^\Lambda\,.
\end{equation}
In particular, when $N=1$ we can put $a_1=1$ and have $\widetilde{\mathfrak{p}}^{\Lambda}=\mathfrak{p}\sqrt{q\hbar}$ as $\hbar\to\infty$ and $\mathfrak{p}\to 0$ so that the latter combination is finite. 

One can see that $\hbar^{n-i}$ in \eqref{eq:aequivspec} will cancel off after plugging into \eqref{eq:E1eigneeRSToda}. As expected, fixed points in the vortex moduli space are parameterized by integers $\lambda_i$ -- vortex numbers.

As it was pointed out by Hanany and Tong in \cite{Hanany:2003hp}, the vortex moduli space $\mathcal{M}^{\text{vort}}_{1,k}$ (the so-called `$\frac{1}{2}$-ADHM' moduli space) forms a Lagrangian submanifold inside the instanton moduli space $\mathcal{M}^{\text{inst}}_{1,k}$. This submanifold is the fixed point locus of a $U(1)$ action on $\mathcal{M}^{\text{inst}}_{1,k}$ which rotates the instantons in a plane. Using the language of Nekrasov's Omega background, we can identify this action with $\mathbb{C}^\times_\hbar$.
\begin{table}[!h]
\begin{center}
\renewcommand\arraystretch{1.2}
\begin{tabular}{|c|c|c|}
\hline
affine q-Toda model& 5d/3d $\mathcal{N}=2$ SYM theory & 3d $\frac{1}{2}$-ADHM theory  \\ 
\hline\hline Coordinates $z_i$  & K\"ahler parameters & K-ring generators $x_i$ \\ 
\hline Eigenfunctions & Defect partition functions & $\frac{1}{2}$-ADHM Coulomb branch vacua  \\
\hline Planck constant $\log q$& equivariant parameter $q$& $\mathbb{C}^\times_q$ acting on $\mathbb{C}$  \\
\hline Affine parameter $\mathfrak{q}$ & 5d dynamical scale $\mathfrak{p}^\Lambda$  & FI coupling $\widetilde{\mathfrak{p}}^{\Lambda}$ \\
\hline  Eigenvalues $\mathscr{E}^{\text{Toda}}_r$ & VEVs of Wilson loop $\langle W_{\Lambda^r \tiny{\yng(1)}}\rangle$ & Chern polynomials $\mathcal{E}^\Lambda_r$ of $\Lambda^r\mathcal{U}$  \\
\hline
\end{tabular}
\vspace{2mm}
\renewcommand\arraystretch{1}
\caption{\small The correspondence table between the closed q-Toda model, its 5d/3d gauge theory description and large-$n$ $\frac{1}{2}$-ADHM quiver description.}
\label{Tab:CorrespToda}
\end{center}
\end{table}

Thus the new duality can be stated as follows. The VEV of a Wilson line in pure $\mathcal{N}=2$ SYM theory with gauge group $U(n)$ with quantized Coulomb branch parameters \eqref{eq:aequivspectoda} in the Nekrasov-Shatashvili limit at large $n$ becomes the VEV of the corresponding flavor-Wilson line of the $U(1)$ 
$3d$ $\mathcal{N}=2$ quiver theory whose Coulomb branch describes the vortex moduli space. 

By examining \tabref{tab:TableFields} we conclude that in the $\hbar\to\infty$ limit adjoint chiral field $B_1$ and anti-chital field $J$ decouple leaving us with only $B_1$ and $I$ ($\chi$ becomes constant due the F-term constraint). This illustrates on the level of the 3d gauge theories how the ADHM model becomes the $\frac{1}{2}$-ADHM model.

\subsection{Generating Function of the $\Lambda$ILW Model}\label{Sec:GenFuncLambdaILW}
Using the \cite{Koroteev:2016,Koroteev:2018a} description of the quantum ILW model we can derive a generating function for the new system, which we call $\Lambda$ILW, by studying the Inozemtsev limit of \eqref{eq:genfuncILW}. 

From \eqref{eq:TripleHeis} we can see that provided the scaling \eqref{eq:pLambdadef} takes place 
the new $(q,\mathfrak{p}^{\Lambda})$-deformed Heisenberg generators have the following form
\begin{equation}
[b_m, b_n] = -\dfrac{1}{m} \dfrac{1-q^{m}}{1-(\mathfrak{p}^{\Lambda})^{m}} \delta_{m+n,0}\,.
\label{eq:TripleHeisToda}
\end{equation}
where we absorbed a divergent factor proportional to $(1-\hbar^{m})^{1/2}$ into the definition of $b_m$. Equivalently, this factor can be absorbed into generating parameter $\xi$. 

Thus we can construct a generating function for $\Lambda$ILW Hamiltonians
\begin{equation}
\mu(\xi) = \text{exp}\left(\sum_{n>0}b_{-n}\xi^n\right) \text{exp}\left(\sum_{n<0}b_{n}\xi^{-n}\right)\,, 
\label{eq:LambdaILW}
\end{equation}
so that
\begin{equation}
\mathcal{H}_{\Lambda\text{ILW}} = [\mu(\xi)]_1\,.
\label{gfLambda}
\end{equation}

Notice the similarity between \eqref{eq:TripleHeisToda} and \eqref{eq:qHeisenb}. It is not accidental as both tRS operators and, as we have just concluded, the qToda operators at large $n$ act naturally on the K-theory of the vortex moduli space $\mathcal{M}^{\text{vort}}$.

\subsection{Geometric Applications}
Therefore we conclude that the spectrum of $\Delta$ILW Hamiltonians is in one-to-one correspondence with the operators of quantum multiplication in $QK_{q}(\text{Hilb}^k(\mathbb{C}))$ by the symmetric powers of the universal bundles.

One can think of a subscheme $\mathscr{Z}_k$ of Hilb$^k[\mathbb{C}^2]$ parametrizing ideals scheme-theoretically supported on $\mathbb{C}\subset\mathbb{C}^2$ (i.e. where the $y$ matrix is identically $0$) is the same as the 1/2 ADHM quiver variety. The complete Hall algebra which acts on $\oplus_k K_{q,\hbar}(\text{Hilb}^k)$ does not preserve the K-theory of this subscheme $\mathscr{Z}_k$, but there is a one-parameter Heisenberg subalgebra inside it that preserves $\oplus_k K_q(\mathscr{Z}_k)$. This Heisenberg subalgebra is the natural analogue of Nakajima's construction.

Our calculations lead to the new results on equivariant K-theory. First, we remind the reader about the following theorem 
\begin{theorem}[\cite{Koroteev:2017}]
\label{todath}
The quantum equivariant K-theory of the complete n-dimensional flag variety is given by
\begin{equation}
QK_{T'}(\mathbb{F}l_n)=\frac{\Complex[\mathfrak{z}_1^{\pm 1},\dots,\mathfrak{z}_n^{\pm 1}; \mathfrak{a}_1^{\pm 1},\dots,\mathfrak{a}_n^{\pm 1}; \mathfrak{p}_1^{\pm 1},\dots,\mathfrak{p}_n^{\pm 1}]}{\left(H^{\text{q-Toda}}_r(\mathfrak{z}_i, \mathfrak{p}_i)= e_r(\mathfrak{a}_1,\dots, \mathfrak{a}_n)\right)}\,,
\end{equation}
where $H^{\text{q-Toda}}_r$ are given by \eqref{eq:tRSRelationsElToda} and $T'$ is the maximal torus of $GL(n)$ with equivariant parameters $\mathfrak{a}_1,\dots, \mathfrak{a}_n$ .
\end{theorem}
Inspired by this result we can prove a theorem about the projective $n\to\infty$ limit of the above ring similar to 
Theorem \ref{Th:EmbeddingTh} when we further specialize the values of $\mathfrak{a}_i$s as in \eqref{eq:aequivspectoda}.

Similarly to $\mathcal{H}_n$ we can define the moduli space of quasimaps to complete $n$-flags (as opposed to the cotangent bundles to those flags earlier in the paper)
\begin{equation}
\mathcal{P}_n:=K_{\textsf{T}}(\textbf{QM}(\mathbb{P}^1,\mathbb{F}l_n))
\label{eq:KthquasimapsFl}
\end{equation}
for extended maximal torus $\textsf{T}'=T'\times \mathbb{C}^\times_q$.

As it was discussed in \cite{Koroteev:2018} the vertex functions (quantum classes) of $\mathcal{P}_n$, under proper normalization, be directly obtained from the vertex functions of $\mathcal{H}_n$. Thus, for a fixed point \textbf{q} of the maximal torus
\begin{equation}
\mathsf{I}_{\textbf{q}} = \lim\limits_{\hbar\to\infty} \mathsf{V}_{\textbf{q}}\,.
\end{equation}

Then the following statement follows:
\begin{theorem}
\label{Th:EmbeddingTh}
For $n>k$ there is the following embedding of Hilbert spaces
\begin{align}
\bigoplus\limits_{l=0}^k K_{q}(\text{Hilb}^l(\mathbb{C})) &\hookrightarrow \mathcal{P}_n
\label{eq:EmbeddingnFl}\\
[\lambda]&\mapsto \mathsf{I}_{\textbf{q}}\,,\notag
\end{align}
where $\mathsf{I}_{\textbf{q}}$ is the K-theory vertex function for some fixed point $\textbf{q}$ of maximal torus $T'$.
The statement also holds in the limit $n\to\infty$
\begin{equation}
\bigoplus\limits_{l=0}^\infty K_{q}(\text{Hilb}^l(\mathbb{C})) \hookrightarrow \mathcal{P}_\infty\,,
\label{eq:Embeddinginf}
\end{equation}
where $\mathcal{P}_\infty$ is defined as a stable limit of $\mathcal{P}_n$ as $n\to\infty$.
\end{theorem}

\section{Towards the Physical Picture}\label{Sec:PhysicsPicture}
In the past sections we have mainly considered formal geometrical ways to perform the Inozemtsev-like scaling limit. Let us attempt to develop more physical interpretation of the dimensional transmutation phenomena in hydrodynamics. 
In QFT a scale anomaly can be thought of as a gravitational phenomenon,
when the cutoff in the theory depends on the external metric. The IR non-perturbative scale
enters into the VEV of the trace of the stress-energy tensor in the ground state $\langle\theta_{\mu \mu}\rangle\neq 0$ resulting in the gap in the spectrum. 
The dimensional transmutation phenomenon is observed in asymptotically free theories in the presence of such scaling anomaly.

\subsection{Vortex Fluid and Scale Anomaly}
Can we recognize these two ingredients of the dimensional transmutation in hydrodynamics? First
let us look for the hydrodynamical theory with the scale anomaly. The proper pattern has been
recently found in \cite{Wiegmann:2019qvx}, namely that is the quantum vortex fluid which describes FQHE and
the  rotating superfluid. The fluid is described by the macroscopic
density of vortices of the same chirality which supports the chiral flow.
It was found that several nontrivial phenomena occur at the quantum level.
The origin of these effects is the UV cutoff introduced by the effective
finite sizes of the vortex cores or, equivalently, the minimal distance 
between the vortices. The cutoff is scale dependent which results in the scale anomaly in the quantum vortex fluid.  

It was demonstrated that the stress tensor and the scale anomaly emerge in the bulk
of the vortex fluid
\beq
T_{zz}= \frac{\hbar \Omega}{12\pi}\left(\partial_z^2 \log n - \frac{1}{2}(\partial_z \log n)^2\right)\,,
\eeq
where $n(r)= \sum_i \delta (r-r_i)$ is the density of vortices and $\Omega$ is angular velocity. This quantum anomaly
modifies the fluid equation, for instance the the
Helmgoltz law gets modified as
\beq
D_t n= \frac{1}{2}\nabla T^{\bar{z}}_z \times \nabla n\,.
\eeq
The vortices are no longer frozen in the bulk fluid flow.

The nontrivial phenomena occur at the edge of the vortex chiral flow
as well \cite{Bogatskiy:2019}. The finite width boundary layer with the different
vortex density gets emerged. 
The vortex density in the bulk is finite, however the quantization
condition for the number of vortices in the boundary layer 
is imposed. Hence we find ourselves in the situation
with the boundary between vortex fluids with two different densities.
The vortex dynamics on the boundary is described by the BO equation.

For the vortex density $n(x,t)$ at the boundary it was shown in \cite{Bogatskiy:2019} that
\beq
\partial_t n(x,t)= -U\partial_x n +\frac{1}{2}\Gamma\,\partial_x\left(n^2 -\bar{d}\,\partial_x n^{H}\right)\,,
\eeq
where $\Gamma$ is circulation of each vortex , $U=\frac{\Gamma}{\sqrt{16\pi l}}$ and $l$
are the mean inter-vortex distance and $\bar{d}$ is the dipole moment.
The width of the boundary layer is determined
dynamically
\beq
h= l^2 \bar{d}\,.
\eeq
We expect to find the ILW equation to describe the boundary vortex dynamics for the generic wave length,
however this point deserves further investigation.

The emerging description of the quantum vortex fluid has many similarities with 
2d quantum gravity \cite{Wiegmann:2019qvx}. 
The density of vortices defines the effective 2d metric in the bulk
\beq
ds^2=n\,dz d\bar{z}\,,
\eeq
and the stress tensor has the Schwartzian form in terms of $n(z,t)$. 
The symmetry of the vortex chiral flow is the S$\text{diff}$ and one can 
also consider the Virasoro algebra of holomorphic reparameterizations 
of the plane. The important $SL(2,\mathbb{R})$ subalgebra of Virasoro and 
S$\text{diff}$ is generated by the quadrupole moments of the vortex density
\beq
L_1=\int z^2 n(z)dz\,,\qquad L_{-1}= \int \bar{z}^2 n(z)dz\,, \qquad L_0=\int z \bar{z}\, n(z)dz\,.
\eeq
At the quantum level the coordinates commute as
\beq
[z,\bar{z}]=i\hbar\,,
\eeq
so that the subalgebra is generated by powers of $z$ and Dunkl operators.

\subsection{Microscopic Arguments}
The dimensional transmutation phenomenon can be recognized in SYM theory 
at the level of single instanton computation. To this aim one evaluates the one-instanton action and the 
regularized determinant of the fluctuations around instanton.
The contributions from nonzero modes get canceled hence the
determinant is solely saturated by zero modes. 
The determinant of the regulators yields the factor
$M^{k}$, where $k$ is the difference between numbers of fermionic and bosonic zero modes.
Together with the instanton 
weight factor $ \exp(\frac{1}{g^2(M)})$
it amounts to the power of $\Lambda$
scale if we take into account the running  coupling constant. 
This trick allowed \cite{Novikov:1983uc} to extract the exact $\beta$-function in the
 $\mathcal{N}=1$ SYM theory. In fact the possibility
to evaluate the exact beta-function in SUSY YM theory at the single
instanton is the simplest manifestation of the resurgence phenomenon in QFT. 
The quantum corrections on the top of the non-trivial saddles are 
related to the calculations in the perturbative sector of the theory.

Having in mind this field theory result we could ask if there is 
some `elementary' object in hydro context  which could indicate
the analogue of dimensional transmutation. To this aim let us
consider the interaction of two vortices which is effectively 
described by the Calogero model. Naively the Hamiltonian of Calogero model  is conformal
invariant and enjoys the familiar spectrum-generating $SO(2,1)$ symmetry with generators 
$J_1,J_2,J_3$ identified with the Hamiltonian, the dilatation
and the special conformal transformation
\beq
J_1= H\,, \qquad J_2= D= tH - \frac{1}{2}(px+xp)\,, 
\qquad J_3 = t^2H - \frac{t}{2}(px+xp) +\frac{1}{2}x^2\,.
\eeq
\beq
H= p^2 + \frac{\nu^2}{x^2}\,.
\eeq
This algebra  is the edge counterpart of the algebra of quadrupoles in the bulk.  
The mass of the regulator $M$ in the gauge theory plays the role 
of the coupling constant $\nu$ in the Calogero model. 

Let us introduce the cutoff $R_0$
corresponding to the minimal distance between two particles.
To some extend it is the counterpart of the effect
of finite-size vortices discussed above.
Careful analysis of the cutoff dependence in two-body 
Calogero model amounts to
a few important observations. It turns out that the scale
symmetry is broken down to the discrete subgroup due
to an anomaly \cite{A_a_os_2003}
\beq
A= -[D,H]+ H \neq 0\,.
\eeq
The anomaly equation reads as
\beq
\frac{d\langle D\rangle}{dt}= \langle H\rangle\,,
\eeq
which is the quantum mechanical counterpart of the QFT anomaly
equation for the dilatation current
\beq
\partial_{\mu}D_{\mu} = \theta_{\mu\mu}\,.
\eeq
The renormalized coupling constant depends now on the
cutoff $\nu(R_0)$ is a peculiar way. 
The RG procedure works as follows. We assume that there is a minimal UV distance
between two particles and determine how the effective coupling depends on the minimal distance.
A bit surprisingly it turns out that the corresponding RG equation for Calogero 
coupling constant admits a limit cycle
that is there is the interplay between the UV cutoff and the IR scale (the review on the 
RG limit cycles aka Efimov phenomena can be found in \cite{BulychevaGorsky:2014}).

In the QFT description we have 
the mass of the regulator $M$ which yields the UV cutoff and the running coupling $g^2(M)$
whose proper combination provides the IR scale $\Lambda$. In the hydrodynamical setting
the interpretation of these two parameters gets reversed in an interesting way. The UV parameter 
$M$ now measures the strength of the interaction between vortices while the gauge coupling $g^2$
yields the geometric scale of the model. Now we search for the $M(g^2)$ dependence which  look a bit unusual from
the QFT viewpoint.

Remark that if we focus at the near $E=0$ part of the spectrum it turns out that there is the tower of the quasi-zero
Efimov-like modes
\beq
E_n=c\exp \left(-\frac{n-n_0}{\nu}\right)
\label{efimov}
\eeq
with some constant $c$.  The energy value
in the elliptic Calogero model corresponds to the VEV 
$u=\left\langle \text{Tr}\phi^2\right\rangle$ in the SYM theory hence the tower of Efimov states
corresponds to the states near $u=0$ in the presence of the $\Omega$-deformation in the NS limit.

We conjecture that the similar picture holds for the many-body Calogero system. We impose
the `finite size' of the vortex as the field dependent UV cutoff which prevents the 
vortices from sitting at one point (falling to the center) and consider how the 
effective Calogero coupling $M$ depends on this UV cutoff. Once again we have the RG cycles
which mean that the UV and IR scales are connected. In the many-body situation
an additional effect occurs. Apart from renormalization of the coupling the vortex lattice 
gets formed and the distance between lattice sites becomes fixed. 

We hope that the above arguments hold in the hydrodynamical limit as well. 
If true then, indeed, we have two physical phenomena which are necessary for 
dimensional transmutation -- the scale anomaly, which unfreezes the vortices from 
the fermionic flow \cite{Wiegmann:2019qvx},
and the running of the coupling constant between the vortices as a function of geometrical scale.
Let us emphasize that these phenomena take place at the boundary between the
bulk layer and the upper layer.

\subsection{An Analogy with the Peierls Model}
We speculate above
that the transition from the ILW hydrodynamics to
Toda hydro limit occurs via the decoupling of the
vortices attached to the atoms of fluid or to fermions
in the FQHE. 
That is the fluxes do no longer flow with the fermionic fluid
but are approximately frozen at fixed positions forming a kind of flux lattice.
To some extend this is the edge counterpart of the observation \cite{Wiegmann:2019qvx}
that the vortices are no longer frozen in the bulk chiral flow at the quantum level.

More technically we assume  that the Baker-Akhiezer function in the finite Calogero
and Toda systems plays the role of the fermionic wave function 
while the vortex degrees of freedom correspond to the coordinates and 
momenta.  In the Calogero case with long-range 
interaction the fermion whose wave function is identified with BA flows coherently with
the fluxes forming a kind of interacting fluid of composite
particles. 
The example of this behavior is provided by the model of 1d superconductivity - the Peierls model.
Some version of the Peierls model admits the exact solution \cite{BRAZOVSKII198240}
being related with the affine Toda model with the clear-cut physical interpretation. 
The Toda chain Lax operator can be considered as a Hamiltonian of the Peierls fermionic system so that the 
spectral parameter plays a role of energy and the classical spectral curve of the Toda system simultaneously provide 
the dispersion relation for fermions. More precisely, the model describes spectrum of quantum electrons interacting 
with the classical Toda potential formed by lattice of heavy ions. The model exhibits superconductivity and it generates a 
non-perturbative scale corresponding to the binding energy of the Cooper pairs. The hydro description can be provided by the
fermions  dressed by phonons. The fermion density is the parameter of the model which
strongly influences the ground state.

There is the link of the Peierls model with the SUSY YM  at finite N. In \cite{Gorsky:1997} it was noted that 
the integrable structure of pure $\mathcal{N}=2$ SYM theory has an interpretation in terms of quasiparticle excitations 
of the Peierls model. It was then argued in \cite{Gorsky:1997} that generation of this 
non-perturbative (BCS) scale can be interpreted on the gauge theory side as dimensional transmutation 
from $\mathcal{N}=2^*$ theory to pure $\mathcal{N}=2$ SYM.

\section{Conclusions}\label{Sec:Conclusions}
In this paper, motivated by the correspondence between large-$N$ $\mathcal{N}=2$ supersymmetric gauge theories 
with adjoint supplemented with BPS defect  and ILW equation in 1d hydrodynamics, we have considered the possible counterpart of the dimensional transmutation
phenomena in hydrodynamics. At the gauge theory side we have elaborated some aspects of the 
instanton counting in the coupled 4d-2d system and in particular the reduction of instanton 
counting in 4d-2d system to the instanton counting in 2d system on the defect. This limit 
corresponds to the open Toda chain in the case of finite number degrees of freedom 
and to the corresponding hydro counterpart at large-$N$. Some new duality is formulated in this case.

We have mentioned a few ways to the  hydrodynamic limit corresponding to the closed Toda chain.
First way concerns the modified pole Ansatz, the second deals with the continuum limit
of the Toda equation of motion in the Hamiltonian formulation supplemented with the
additional chiral constraint while the third way involves the Inozemtsev-like limit directly
in  the periodic ILW equation. It is not completely clear if these procedures are 
equivalent due to the possible non-commutativity of the limits. We plan to clarify these
issues in the further study.

We conjecture that the chiral flow in the vortex fluids provides the proper playground
for the dimensional transmutation phenomena. The BO equation emerges naturally 
at the boundary between the bulk of the vortex fluid and the boundary layer. Inspired
by the analogy with the Peierls model we assume that in the Toda-like hydro 
microscopically the flow occurs at the top of the weakly fluctuating vortex lattice
generated at the boundary. The very phenomenon of dimensional transmutation
presumably is interpreted as the geometric renormalization of the coupling
between vortices when naively coupling tends to infinity. The emerging 
finite coupling is analogue of the $\Lambda$-like scale. The key 
point responsible for this phenomena is the scale anomaly 
in the vortex fluid. Certainly this
interpretation deserves for further clarification.

There are many questions to be elaborated, just mention a few. 
In the field theory context we can derive pure YM theory in two ways-
starting with the theory with adjoint or fundamental matter and decouple
it yielding the non-perturbative scale via dimensional transmutation. 
In our study we started with the theory with adjoint matter  which
has ILW hydro counterpart however it would be interesting to find
the hydro description of the theory with the fundamental matter
and perform the limiting procedure to Toda theory in this case as well.
The cascade of the phase transitions has been found in $\mathcal{N}=2^*$ at large-$N$ \cite{Russo_2013}
if the 4d instantons are switched off. It would be interesting to investigate
if there are some traces of these phase transitions which have nontrivial 
holographic description if the surface defect is added in the hydrodynamic
description. Presumably these phase transition could correspond to the
particular solutions to the BO equation. 

It would
be important to understand the exact role of the Gaiotto-Whittaker state 
in $W_{1+\infty}$ algebra in the Toda-like hydrodynamics which would explain the
algebraic interpretation of the emerging non-perturbative parameter.
According to AGT the wave function of the quantum 
periodic Toda chain for any N 
is related to the matrix element over the state  $\Psi$ which 
is the  Whittaker vector for Virasoro algebra  and provides
the irregular Liouville conformal block \cite{Gaiotto:2009ma}
\beq
L_1 |\Psi\rangle = \Lambda |\Psi\rangle \qquad \langle\Psi|L_{-1}= \langle\Psi|\Lambda
\eeq
where $L_1, L_{-1}$ are the Virasoro generators. Since we are hunting 
for the hydrodynamical interpretation of $\Lambda$ 
it is natural to assume that the same 
Gaiotto state is involved into  the quantum hydrodynamics in the Toda limit. Hence
the non-perturbative scale probably plays the role of the intrinsic momentum involved
in the chiral quantum fluid.

One more question concerns the possible relation with the 
particular limit of the torus knot invariants. The point is that
the equation of motion for the finite number of interacting
vortices in the bulk of the vortex fluid are written in terms of Dunkl-like operators. 
On the other hand weighted multiplicities 
of the eigenfunctions of the Dunkl are related to the invariants 
of the torus knots at the rational coupling constant(see, i.e. \cite{Bulycheva:2014}).
In the hydro limit the coupling constant between the vortices naively
tends to infinity which means that $T_{\infty,n}$ torus knots are relevant.
In this stable limit of the torus knots the new algebraic structures
get emerged \cite{gor} hence it would be interesting if the 
relation with the torus knot invariants survives at the edge of the
vortex fluid.

\bibliography{cpn1}
\end{document}